\documentclass[sigconf,authorversion]{acmart}
\pdfoutput=1
\usepackage[utf8x]{inputenc} 
\usepackage[T2A]{fontenc} 
\DeclareUnicodeCharacter{0430}{\-} 
\usepackage{url}
\usepackage{amsmath,amsfonts}
\usepackage{algorithmic}
\usepackage{graphicx}
\usepackage{xcolor}
\usepackage{float}
\usepackage{subcaption}
\usepackage{cleveref}
\usepackage{siunitx}
\long\def\comment#1{}
\newcommand{\vvv}{\vspace*{0.0cm}}
\AtBeginDocument{%
  \providecommand\BibTeX{{%
    \normalfont B\kern-0.5em{\scshape i\kern-0.25em b}\kern-0.8em\TeX}}}

\copyrightyear{2019}
\acmYear{2019}
\setcopyright{rightsretained}

\begin{document}

\title{Let Your Camera See for You: A Novel Two-Factor Authentication Method against Real-Time Phishing Attacks}

\author{Yuanyi Sun}
\affiliation{%
  \institution{Penn State University}
  \city{University Park, PA 16802}
  \country{USA}}
\email{yus160@psu.edu}

\author{Sencun Zhu}
\affiliation{%
  \institution{Penn State University}
  \city{University Park, PA 16802}
  \country{USA}}
\email{sxz16@psu.edu}

\author{Yao Zhao}
\affiliation{%
  \institution{Shape Security}
  \city{Santa Clara, CA}
  \country{USA}}
\email{yzhao@shapesecurity.com}

\author{Pengfei Sun}
\affiliation{%
  \institution{Shape Security}
  \city{Santa Clara, CA}
  \country{USA}}
\email{psun@shapesecurity.com}









\begin{abstract}
Today, two-factor authentication (2FA) is a widely implemented mechanism to counter phishing attacks. Although much effort has been investigated in 2FA, most 2FA systems are still vulnerable to carefully designed phishing attacks, and some even request special hardware, which limits their wide deployment. Recently, real-time phishing (RTP) has made the situation even worse because an adversary can effortlessly establish a phishing website 
without any background of the web page design technique. Traditional 2FA can be easily bypassed by such RTP attacks. 
In this work, we propose a novel 2FA system 
to counter RTP attacks. The main idea  
is to request a user to take a photo of the web browser with the domain name in the address bar as the 2nd authentication factor.     
The web server side extracts the domain name information based on Optical Character Recognition (OCR), and then determines if the user is visiting this website or a fake one, thus defeating the RTP attacks where an adversary must set up a fake website with a different domain.
We prototyped our system and evaluated its performance in various environments. The results showed that PhotoAuth is an effective technique with good scalability. We also showed that compared to other 2FA systems, PhotoAuth has several advantages, especially no special hardware or software support is needed on the client side except a phone, making it readily deployable. 
\end{abstract}

\maketitle

\section{INTRODUCTION}

Phishing is a fraudulent attempt to obtain sensitive information, such as usernames, passwords, credit card numbers of users. 
Specifically, an adversary may build a fake website that has a similar domain and web page layout with the real one. Then he spreads the fake domain link (FDL) to benign users (e.g,  through SMS messages, emails, social websites, etc). Some users may treat a FDL as a genuine one, follow the FDL and get directed to the fake website. After entering their credentials to log into the fake website, users leak their sensitive information to the adversary.

Two-factor authentication (2FA) is a user authentication technique which requires end users hold at least two types of information that can confirm his claimed identities, based on something they know, something they have, or something they are. With 2FA, even though users' passwords are stolen, their accounts are still protected as long as the attacker is unable to obtain the second authentication factor. Therefore, 2FA has seen increasing deployment recently, with 53\% surveyed users adopting it for account protection in 2019~\cite{2FAhit}.  

While 2FA can significantly improve account security,  
the arms race with phishing has never stopped. 
Traditional 2FA is still vulnerable to phishing 
because a deceived user may input their second factor  information (e.g., PINs received through emails, SMS) into the fake website, which defeats the second layer of protection. 
Moreover, in the past an adversary had to manually design the web page layout to mimick the real one, which is time-consuming for the adversary.
Recently, the new real-time phishing (RTP) tools, like ``Evilginx" \cite{Evilginx}, have made the situation even worse. Now, an adversary only has to download the tool and run it with proper configuration to automatically replicate from the real website. In other words, RTP tools have significantly lowered the technical barriers for adversaries to launch more powerful phishing attacks.

Several methods have been proposed to detect the traditional phishing by measuring the similarity of web page elements (e.g, image size, position) \cite{medvet2008visual} or tree structures of two websites \cite{rosiello2007layout}. Unfortunately, such methods would not work in RTP, because the fake website keeps replicating its content from the real website through reverse proxy.  
For the same reason, human interactive proof based methods, which rely on human users to determine if the dynamically generated image surrounding the login window matches the locally generated image by the browser extension~\cite{dhamija2005battle,dhamija2005phish}, will fail. Finally, phishing detection algorithms~\cite{aggarwal2012phishari} based on suspicious domains will fall short because an adversary can change its domains frequently.

Recently, new 2FA systems have been introduced and deployed, such as Duo Push~\cite{Duo_Push}, U2F~\cite{U2F}. The research community has also proposed novel proof-of-concept 2FA systems~\cite{karapanos2015sound,czeskis2012strengthening,shirvanian2014two}. However, most of them require special devices (or hardware configurations)~\cite{U2F,karapanos2015sound}, or some are still vulnerable to RTP attacks~\cite{Duo_Push,karapanos2015sound}. 


In this work, focusing on defeating the advanced RTP attacks, 
we propose a new 2FA system called \textit{PhotoAuth}. Here, after a user passes the first factor (e.g., password) authentication, the web server will require the user to take a photo of the web browser with domain name in the address bar as the 2nd authentication factor. The phone automatically uploads the photo to the web server through a web app invoked in the browser of the phone.    
The web server side extracts the domain name information based on Optical Character Recognition (OCR), and then determines if the user is visiting this website or a fake one, thus defeating the RTP attacks where an adversary must set up a fake website with a different domain.



Compared with many other 2FA methods, PhotoAuth has several advantages. First, PhotoAuth can counter RTP attacks that traditional 2FA cannot handle. Second, PhotoAuth does not need any special device other than smartphones that are commonly used. 
No additional extension/plugin or Bluetooth is required for the browser, so users can even log into the web server securely on a public computer through PhotoAuth. 

This work presents the design of PhotoAuth on both the phone side and web server sides. To build an efficient and attack-resilient PhotoAuth, we 
train a deep learning model for address bar detection based on transfer learning, and combine it with OCR output to extract the domain names correctly. 
We tested the accuracy and efficiency of PhotoAuth under different environments. 
The results showed that PhotoAuth is an effective technique with good scalability, and it is readily deployable. 

In summary, we make the following contributions:

\begin{itemize}
\item We propose PhotoAuth, a novel 2FA mechanism based on browser photos to counter real-time phishing attacks, homographic phishing attacks, and domain injection attacks. 
\item Neither the computer nor the phone needs to pre-install any software or apps to finish 2FA, which differentiates it from many other 2FA systems. Especially, no Bluetooth or special devices are needed in PhotoAuth and it is compatible with most legacy devices.
\item An address bar content recognition mechanism is proposed to counter adversaries from making fake domains at the titles or web content. We labeled and trained the first address bar detection model with 16,454 items.
\end{itemize}

Note that PhotoAuth is designed and implemented for the case when a user uses a PC browser to log into a website with his phone as his 2nd factor device. We understand that nowadays many people also use the same phones for website login. We will provide a variation of our method to defeat the RTP attacks, as elaborated in Appendix. \textit{The main body of our presentation will focus on the first case which involves both PC and phone.}


\section{PRELIMINARIES}

\subsection{Real-time Phishing Workflow}
\begin{figure}[!t]
\centering
\includegraphics[scale=0.45]{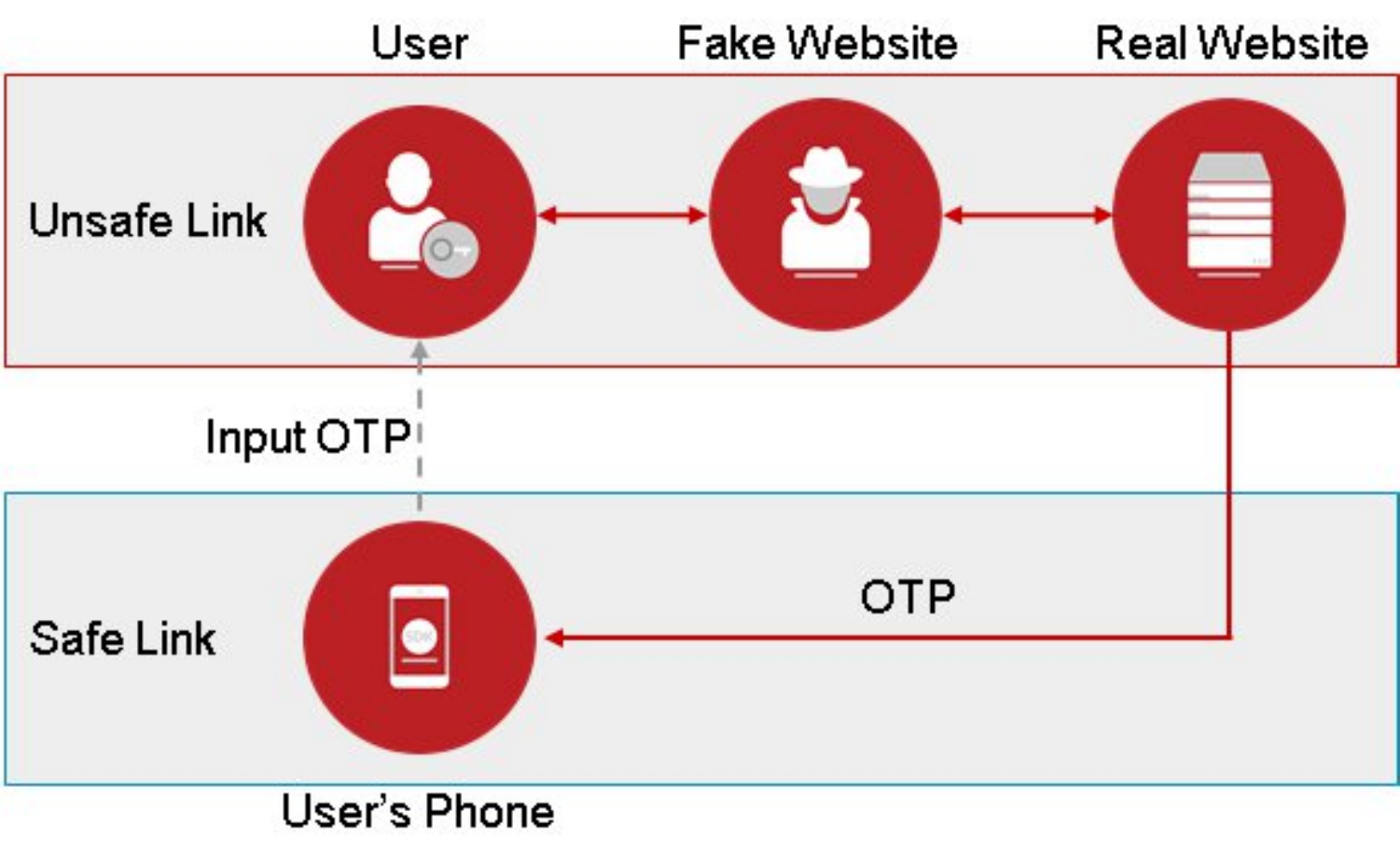}
\caption{Real-time Phishing (RTP) Workflow with OTP}
\label{fig:RTP_workflow}
\end{figure}

Figure~\ref{fig:RTP_workflow} shows how a real-time phishing (RTP) works, where the adversary is in the middle of the benign user and the real website. The detailed steps are as follows.

\begin{itemize}
    \item Step 1: The adversary sets up a fake website (microsoft1.com), which replicates a target website (e.g, microsoft.com), with a mature RTP tool (e.g, Evilginx~\cite{Evilginx}). With proper settings, the RTP tool can establish the fake website automatically and make it a man-in-the-middle web proxy for microsoft.com. Then the adversary distributes the url of the fake website to users through phishing channels.
    
    \item Step 2: A user (referred to as \emph{Bob}) does not pay close attention to the domain name and treats the fake website as the real one because of the same web-page layout. In this example, Bob inputs his Microsoft user name and password to microsoft1.com, the fake Microsoft phishing site.

\item  Step 3: The adversary gets Bob's credentials after Bob submits them to the fake website. Then the RTP tool opens a new session to access the real Microsoft website, and enters Bob's credentials to login. Now the adversary impersonates Bob on the Microsoft website. During the process, the RTP tool automatically modifies appropriate message fields when relaying them between the benign user and the real website so that neither the user or the web server notices the difference from normal use cases. 

\item Step 4: To verify the login, the Microsoft website sends a one time password (OTP) to Bob's phone.

\item Step 5: Bob gets the OTP from the Microsoft website and inputs it into the fake website for authentication.

\item Step 6: The adversary gets Bob's OTP and inputs it into his own login session to the Microsoft website as Bob. Finally, he successively logs into Bob's account. 
To the Microsoft website, it gets a valid login request from the user Bob without knowing the existence of an adversary in the middle.

\end{itemize}

\subsection{Security Model}

On the server side, we assume it has implemented a 2FA mechanism. The server first verifies the username and password provided by the user. If correct, the server challenges the user with something only the user has through his phone to finish the second authentication. 
The target users of our new 2FA system are ordinary people like users of other 2FA systems. They know basic things about web apps like clicking the link to open with a browser. They also have the knowledge of 2FA.  

Certainly, no 2FA system is attack-proof if we assume the second factor can also be compromised.  
In our system, we assume neither the user's PC browser nor his smartphone is compromised by the adversary. The link (referred to as \textit{phone link}) between user's smartphone and the website is also secure from interception (e.g., man-in-the-middle attack \cite{Man-in-the-middle_attack}). 
We only consider the RTP attack alone. That is, we do not consider phishing mixed with other attacks like DNS spoofing \cite{DNS_spoofing}, Domain hijacking \cite{Domain_hijacking}, browser hijacking \cite{Browser_hijacking}, or system compromises.

\subsection{Design Goals}
We have the following design goals for our new 2FA system. 

\noindent\textbf{High Compatibility}
The system should be compatible with the traditional 2FA system to support most legacy devices. The traditional 2FA system can be upgraded to the new 2FA system easily without changing much on the 2FA workflow. 
\noindent\textbf{High Usability}
For pervasive deployment of our 2FA system, no special hardware other than smartphones will be required. To support 2FA, we will not require the installation of browser plugin/extension; otherwise, users of a public computer (e.g, a public library computer) will not be able to use it. Moreover, some non-tech-savvy users may not know how to install browser plugins/extensions.   

During the 2FA, the system will challenge a user for something beyond username and password. The instructions of the challenge should be as simple and intuitive as possible, so that a user does not need any special background knowledge in a specific area to answer the challenge. Verbose and complicated challenges that are time consuming to answer or hard to understand will easily kill usability, so they should be avoided in our design. 

\noindent\textbf{High Accuracy} The system should provide high accuracy for authentication. In other words, it should incur very low false positive rate and low false negative rate at the same time. False positives happen when benign users failed to pass the 2FA even though they followed the procedure; false negatives happen when the adversary was able to pass the 2FA while impersonating a benign user. As our system assumes the first (i.e., password-based) authentication factor is unreliable, here the high accuracy requirement is only upon our second authentication factor, which should be very difficult or not possible for the adversary to forge. 

\comment{
\subsection{Real-world Solutions} \label{sec:whyfail}
With the above design goals in mind, next we explain why the existing solutions are not sufficient to fully meet our requirements.

To address the password problem, in 2017 the FIDO Alliance proposed a Universal 2nd Factor (U2F) protocol~\cite{U2F}, where end users carry a single U2F device which works with any relying party supporting the protocol. Later, the FIDO Alliance proposed FIDO 2~\cite{Fido2}, by integrating its  Client-to-Authenticator Protocol (CTAP) with W3C's Web Authentication (WebAuthn). 
Users may log into internet accounts using their preferred devices. Web services and apps can turn on this functionality via biometrics, mobile devices and/or FIDO security keys.

While U2F/FIDO2, based on public-key cryptography, can counter phishing attacks very well and look promising, it may take a long time to be widely deployed because of 
several possible factors such as 
cost and usability issues~\cite{slow2adopt}. 
For instance, 
U2F devices are not free, commonly ranging from 20 to 60 dollars, hence a non-trivial cost overhead for either end users or companies. 
Other use options may require pairing between phone and PC, or BlueTooth or NFC. Last, while the concepts and procedure for deploying U2F/FIDO2 could still look complicated to some non-tech-savvy users because of the needed registration, installation or configuration. 
  

Google 2-Step Verification \cite{Google_2-Step_Verification} is a phone application to generate a time-based one-time code for the user in every 30 seconds. No network connection is required between the app and the server. The mechanism requires the user to manually enter the one-time password into the browser. 
Duo Push \cite{Duo_Push} is also a phone application that receives push information when the user sends a request in the browser login page. Duo Push has a user interface with two buttons. The user can decide to approve the authentication or not by taping the corresponding button. 
Unfortunately, Google 2-Step Verification and Duo Push are vulnerable to both the traditional phishing attack and the RTP attacks. 
In the 2nd factor authentication phase, the adversary can deceive the user to pass his one-time password or press the ``Approve'' button in the above two applications. The user may think the 2FA is for himself to authenticate to the web site, but the truth is that the 2FA is for the adversary to get authentication. 

Recently Google released a new software-based 2FA tool leveraging phone's built-in security key \cite{Google_phone's_built-in_security_key}. It requires pre-installed phone app to generate the key, special built-in browser support, and Bluetooth (or NFC) to establish a secure channel between the computer and the phone, such requirements could restrict its usability.  
}

\section{SYSTEM ARCHITECTURE AND DESIGN}

\subsection{Design Considerations}
To defeat the RTP attack, the first step is to detect it through distinguishing benign users from adversaries. Generally, there are three aspects for making such a distinguish.

The first aspect is based on the fact that benign users and adversaries have different IP addresses. If the web server has obtained an authenticated list of IP addresses used by each user (e.g., through explicit/implicit registration or valid historical use), it may base the authentication on IP address.  However, as roaming is so common these days with mobile devices (e.g., with laptops and smartphones), many users do not have a fixed IP address. As such, this approach will not work well in reality, although IP address information may still be leveraged to assist in one way or another. 

The second aspect is that the real web server and the fake one have different SSL/TLS certificates.  
However, a browser only checks if a website is not its block list and has a valid certificate. It cannot determine if it is a phishing website or not. If a website has a valid certificate and hence encrypts the traffic between the users and the website, the browser will show a green locker icon. When users see such a green locker icon, they may believe they are visiting a trusted website and no one can eavesdrop the  communication. What users do not know is that the RTP tool like Evilginx is able to automatically
obtain a valid SSL/TLS certificate for free (e.g., from Let'sEncrypt) and provide responses to ACME challenges \cite{Next_Generation_of_Phishing_2FA_Tokens}, using its in-built HTTP server. 

The third aspect is that the real website and the fake one have different domain names. While the adversary cannot register the same domain name that is owned by the real website, he can register a very similar one (microsoft1.com) or seemingly valid one (e.g., microsoft.com.jp) to confuse the users. 

As the RTP attack is getting so advanced, in our design we do not rely on end users to detect the attack by themselves. That is, we do not assume users are able to detect a phishing website based on domain names or browser green lock icons. Instead, it is the job of the web server's side to distinguish benign users from RTP adversaries. 
In a nutshell, our system, like FIDO2~\cite{Fido2}, leverages the third aspect, i.e., domain names, to distinguish benign users and the adversaries, but on the server's side, with a software 2FA mechanism involving smartphones only. No hardware device or pre-install app is needed. Technically speaking, the main task is to deliver (the domain name part of) the website URL in the browser address bar to the authenticated web server in a convenient and secure way. If usability is not a concern, there could be many ways to achieve this goal. For example, with built-in browser support customized 
for U2F, the browser passes the URL to the local mounted USB U2F device for signing and then transfers it to the web server for verification. 



\comment{
\begin{figure}[htbp!]
    \centering
      \begin{subfigure}[b]{0.45\textwidth}
        \includegraphics[width=\textwidth]{image/RTP_workflow2.pdf}
          \caption{Real-time Phishing (RTP) Workflow}
          \label{fig:RTP_workflow2}
      \end{subfigure}
      \hfill
      \begin{subfigure}[b]{0.45\textwidth}
        \includegraphics[width=\textwidth]{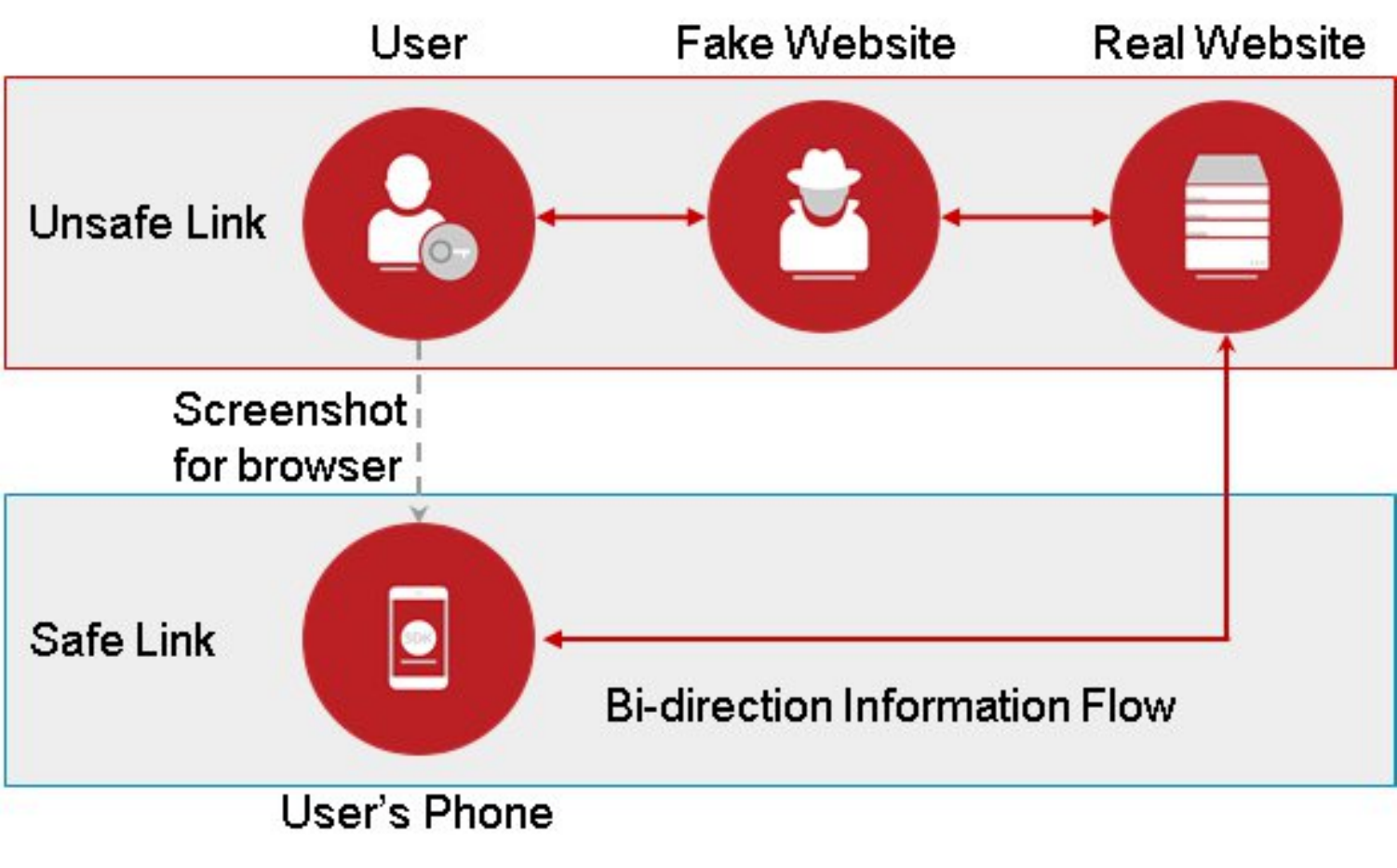}
          \caption{PhotoAuth Workflow}
           \label{fig:PhotoAuth_workflow}
      \end{subfigure}
\caption{Real-time phishing workflow and PhotoAuth workflow comparison.}
\label{fig:one_hash_represent}
\end{figure}
}

\begin{figure}[!t]
\centering
\includegraphics[scale=0.45]{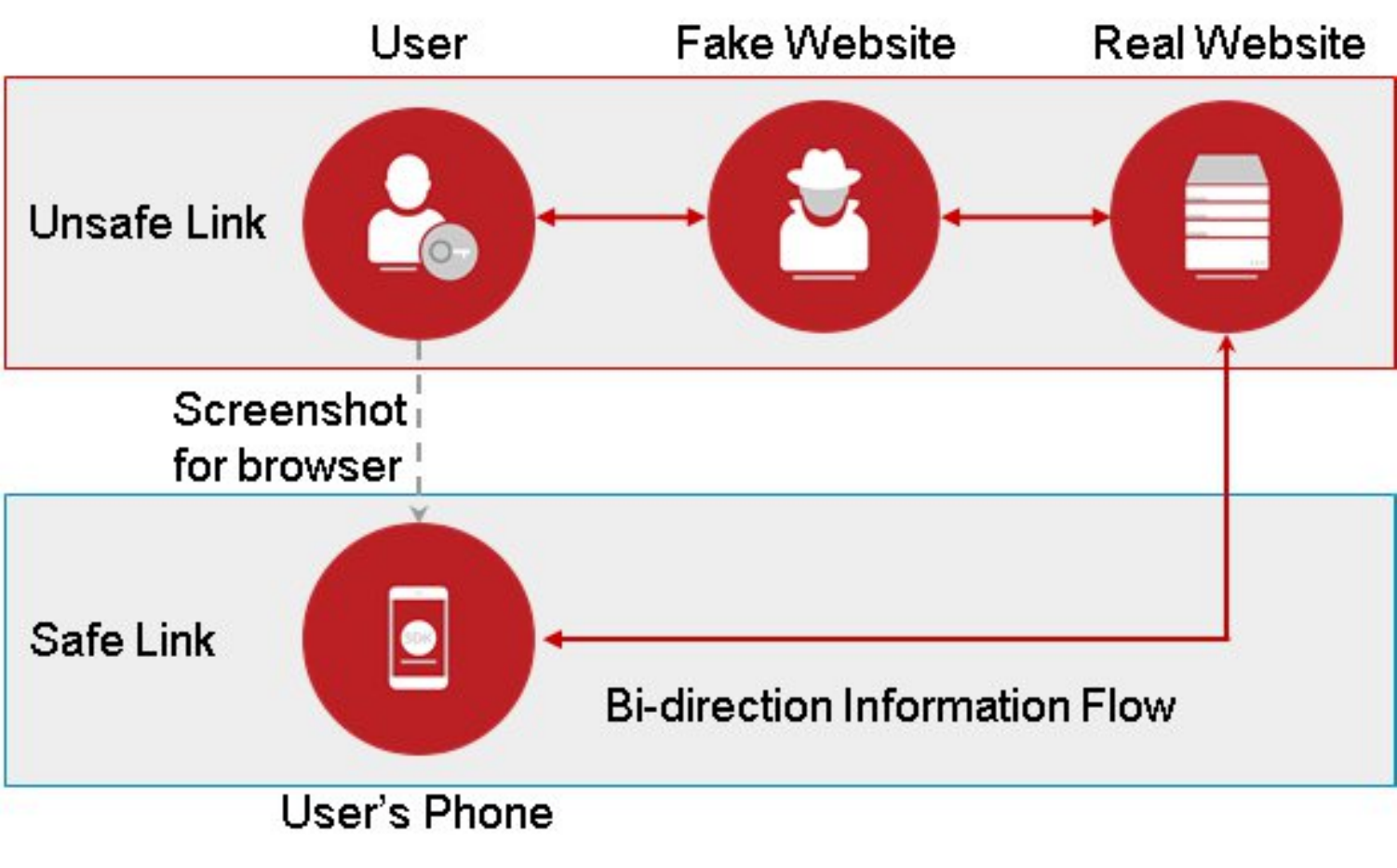}
\caption{Real-time Phishing (RTP) Workflow with PhotoAuth}
\label{fig:photoauthworkflow}
\vspace{-1cm}
\end{figure}

\subsection{System Overview}
\label{systemoverview}
\comment{
\begin{figure*}[!t]
\centering
\includegraphics[scale=0.45]{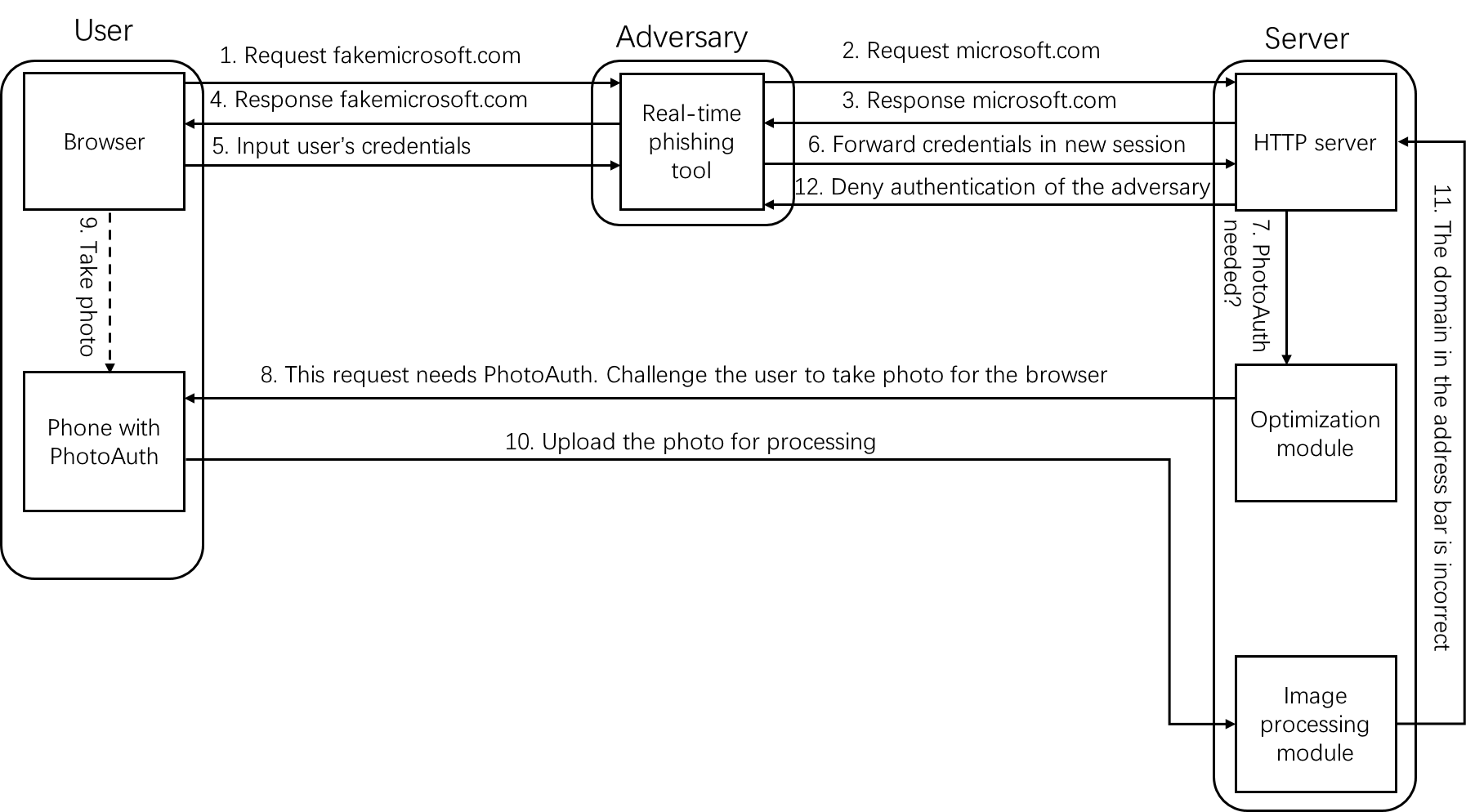}
\caption{System overview}
\label{fig:System_overview}
\end{figure*}
}

\begin{figure*}
\centering
\includegraphics[scale=0.4]{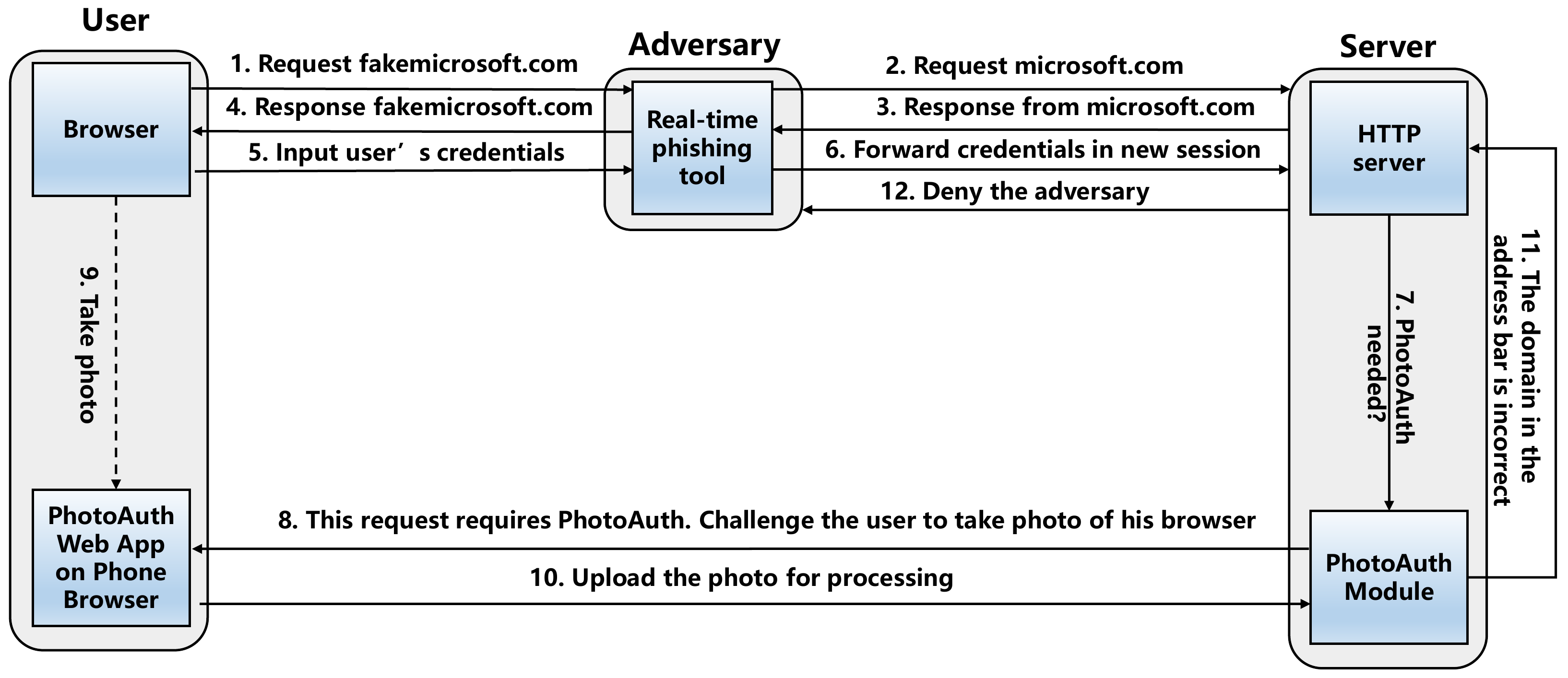}
\caption{System overview}
\label{fig:System_overview}
\end{figure*}

\noindent \textbf{System Workflow:} Figure \ref{fig:photoauthworkflow} shows the high-level workflow of our system, \textit{PhotoAuth}, under RTP attacks. When a benign user visits a RTP-based phishing website through his PC browser and inputs his credentials, the phishing tool opens a new session to authenticate to the real server as the benign user. The real server determines that this new session needs to pass 2FA, so it requests the user to take a photo of his PC browser and send it back through a PhotoAuth web app (a JavaScript-based server-side webpage loaded by the phone's browser). Indeed, what truly matters is the domain name; in other words, our scheme does not require a full URL and the browser size does not matter either except it is too small to show the domain name part. 
Based on the domain name the server tells whether the user is accessing the real server itself or a fake website.

Comparing PhotoAuth to one-time password (OTP) under RTP attacks (shown in  Figure~\ref{fig:RTP_workflow}), there are two major differences. First, in OTP, the phone link (i.e., the one between user's phone and server, which is considered safe) is uni-directional, whereas in PhotoAuth, it is bidirectional. Second, in OTP, a user provides the OTP information to his PC browser, which then passes it to the website through an \textit{unsafe} link, whereas in PhotoAuth, the PhotoAuth web app gets the real information from user's PC browser, and then passes it to the web server through the \textit{safe} phone link. 
Essentially, in PhotoAuth, the trust model is extended from the server itself to user's phone that runs the PhotoAuth web app (i.e., a web page loaded from the server), and the phone takes the role of the authenticator for the website. All sensitive information flows through the safe link rather than the unsafe link. 
Figure~\ref{fig:System_overview} shows the system architecture and detailed workflow of PhotoAuth. On the user side, there is a \textit{PhotoAuth web app} running on the phone browser and triggered by clicking a link pushed from the server. This web app is not a standalone app to be installed on the phone, but a server side web page implemented with HTML5+CSS+Javascript.   
On the server side, there is a \textit{PhotoAuth module}. 
In the workflow, steps 1-6 are similar to those steps in RTP workflow (Section 2.1), so we will not re-introduce them except Step 3. In Step 3, the web server's response includes a web cookie, according to the HTTP protocol and today's web security requirement. 

In Step 7, the real server consults the PhotoAuth module to decide whether a login request requires the PhotoAuth 2FA. 
If it is required, the server sends a notification message containing a unique link to the phone (e.g., through push, SMS, email, etc, as long as it is considered a safe link) (Step 8), which, after being clicked on the phone, invokes the PhotoAuth web app in the user's phone.
For example, when a user (say Bob) logs into the web server (say microsoft.com) with his PC, the server keeps its PC session id. The server then sends a notification message to his phone, which contains a short link like ``microsoft.com/c/6895272031''. Each link sent by the server is unique, where the 10-digit number is randomly generated and stored together with Bob's PC session id for future lookup. In this way, this random number is linked for this specific login. 
Now, when Bob clicks the short link on his phone, his phone browser will load the web page ``microsoft.com/c/6895272031'', which hosts a simple web app with JavaScript code to start the following task. This is to prevent others from submitting responses while impersonating the user. Under a RTP attack, the attacker will not know the short link for submitting, so this method can easily defeat the attack. Moreover, we can increase the length of the random number to defeat brute force attacks. 

In Steps 9 and 10, the PhotoAuth web app 
guides the user to take a photo and automatically upload it to the server for processing. 
In Step 11, the PhotoAuth module extracts from the photo the domain name in the address bar of the PC browser based on deep learning and OCR techniques. 
If the domain name does not match its own one, the user must be visiting a fake website and the authentication request must be from an adversary, so it notifies the the server to deny the authentication request.
In the rare case of a false positive, 
caused by the poor quality of the photo, the user may retry the login process and provide a different photo.  

\comment{
\noindent \textbf{Multiple Browsers in Photo:} But if a user first time uses our system, the user may accidentally place several browsers on the screen and capture them all as a photo to send to the server for authentication. In this case, there are two possible solutions. 

The first solution: after address bar detection, the server would know how many address bars detected in the photo. If more than one address bars detected and they all have no overlap, that means there are more than one independent browsers insight. The server would respond to the user's phone a message that multiple browsers are detected, make the browser you need to authenticate to the full screen and try again. With this design, the user can send a photo with just one browser inside. If the domain is correct, the server authorizes the user. If not, denied the user.

The second solution: when a user sends a request through an adversary to the real server, the real server returns a special icon (one-time password inside the icon, OR code also works) located on the top left of the webpage. The server checks that the user may under phishing and challenges the user to take a photo of the browser. The user places multiple browsers on the screen and takes a photo of them (A browser that has special icon is the one that sent the request). The server gets the photo from the user's phone. This time, the server not only detect address bar location but also detect special icon location. The server use contour detection to check the browser that has the special icon is not overlapped by other browsers. If there is overlap, the server feedback the user to make sure the browser has the special icon on the foreground. If no overlap detected, do OCR on both the address bar and the special icon to extract the domain and the icon one-time password. Finally, the server checks the one-time password is binding the user's session correctly and the domain is also correct. If both correct, the server authorizes the user. If not, denied the user.

With the above two approaches, we solve the problem of multiple browsers inside the photo. As long as the server gets the wrong domain, the server can deny the user and alert the user of the threat of phishing. If the user cannot do the authentication, the user can refer to alternative authentication mechanisms (e.g, SMS one-time password).
}

\comment{
\subsection{Performance Optimization Module}
In the real-world cases, users who have been hijacked by the RTP attacks should be the minority among all users of a system. That is, most user logins (authentication requests) are normal. Now, if PhotoAuth requires all users to take photos of their browsers for authentication, that will incur much burden for the server. Below we present several strategies to optimize for performance.


Figure \ref{fig:system_optimization} shows the flowchart for our performance optimization mechanisms. 
Take Microsoft as an example again. When the Microsoft web server gets Bob's authentication request (the top rectangle symbol in the figure), it first checks the cookie of the request to see whether Bob has been authenticated previously. If yes, the server knows this is an ongoing session, so it authorizes the authentication request directly (The flow goes to the right branch in Figure \ref{fig:system_optimization} ). Requests with invalid cookies may result in the ``400 Bad Request" error (we ignore this case in the figure).  

If the request has no cookie, it is a new session. In this case, the server checks its record for Bob's preferred 2FA method. 

If the preference is to receive an SMS/Push message (preference email has same flow), he will receive an SMS/Push message through his phone, which contains a short link like microsoft.com/c/689527. Here the 6-digit number is Bob's session id. 
Microsoft server obtains the cookie of Bob's authentication request on Bob's phone when Bob clicks the short link on his phone. It then compares this cookie with the one in the previous authentication request. 



If they are the same, then it is the case where Bob is using the same browser to login and to handle SMS/Push short link, it is a legitimate request from Bob. Hence, it will be authorized by the server. SMS/Push messages must be received by phone. That also means Bob is using phone browser to login and click the short link. Otherwise, no matter that it is because Bob opens the short link in a different browser or there exists a RTP attack, the server proceeds with PhotoAuth 2FA. If the cookie is different, it has three cases under RTP phishing, PC sends login but clicks the short link on the phone, the phone sends login but clicks the short link with another phone browser. PhotoAuth can handle the first two cases. For the third one, the server can pop up a message to remind the user that the login request is from the phone, the user should use the same phone browser to open the short link.

With the above performance optimization, only a small percent of users need to go through the PhotoAuth 2FA. 
}

\subsection{User Side Design}

On the user's phone, there are generally three ways to receive notifications from the server, e.g.,, through push message, SMS, or email link.
When the user clicks the notification on the phone, the default browser will be launched (if not yet), which in turn loads the PhotoAuth web app (from the server) to handle the server challenge. 
After that, the user only needs to click one button to take a photo and then upload it in the background. 

In particular, image pre-processing resizes the raw photo image into a smaller and lower resolution image, and further converts the RGB image into a gray-scale one. By leveraging the computational resource of edge devices, we can reduce the bandwidth overhead and hence increase the scalability and throughput of the entire system without sacrificing the accuracy.  Detail of compression ratio and bandwidth saving can be found in the evaluation section. 

\subsection{Server Side Design}
\label{sec:server}
After the server receives the preprocessed photo from a user's phone, it should accurately extract the domain name in the browser address bar.
There are two design options here. First (O1), it may extract all the text contents from the entire photo based on Optical Character Recognition (OCR), and then predict which texts are from the correct domain name (that is, located inside the browser address bar). Second (O2), it may first predict the address bar from the entire photo, and then apply OCR to extract the texts from the predicted address bar only. 

\begin{figure}[htbp!]

    \centering
      \begin{subfigure}[b]{0.45\textwidth}
        \includegraphics[width=\textwidth]{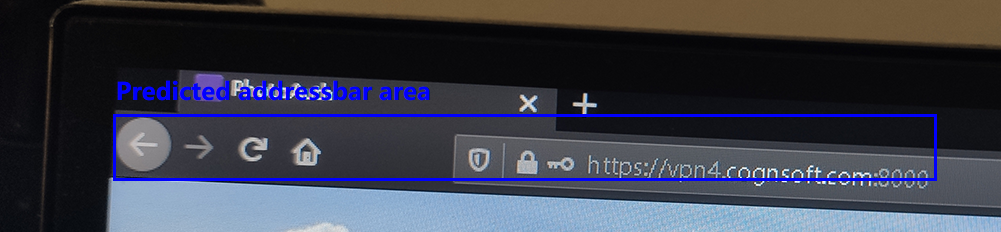}
          \caption{O2 fail example}
          \label{fig:lean}
          \vspace{0.5cm}
      \end{subfigure}
      \vspace{-0.2cm}
      \begin{subfigure}[b]{0.45\textwidth}
        \includegraphics[width=\textwidth]{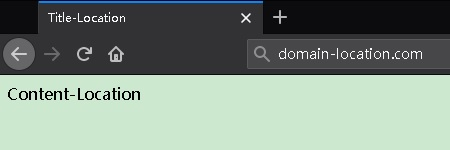}
          \caption{Photograph of a simple webpage (upper-left corner)}
          \label{fig:ocr_example}
          \vspace{0.5cm}
      \end{subfigure}
      \vspace{-0.2cm}
      \begin{subfigure}[b]{0.4\textwidth}
        \includegraphics[width=\textwidth]{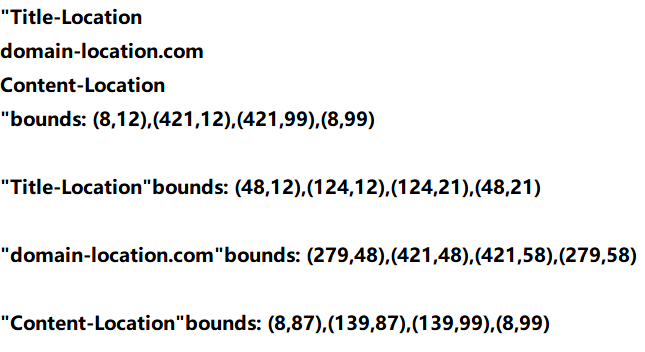}
          \caption{OCR result for the above example}
          \label{fig:ocr_example_result}
      \end{subfigure}
\caption{An Example for Browser OCR}
\label{fig:ocr}
\vvv
\end{figure}

Which of the two options is better? It depends on two factors. O2 sounds more appealing than O1 because it can reduce the workload of OCR by only focusing on a very small region, and the output of OCR is precisely the domain name. In practice, however, this may not be the case, because the time complexity of OCR is not linear to the area of the region. Indeed, the saved time could be very limited. Second, whatever technique we will use, address bar prediction is not going to be perfect. As a result, the bounding box for a predicted address bar may not cover all its texts (e.g., top or bottom side of the texts may be cut off), causing the OCR to fail to output the correct domain name. Figure~\ref{fig:lean} shows such an example.

On the other hand, O1 has a performance advantage over O2, because parallelization is possible. Specifically, for each received photo, the server can first make a copy, and then perform OCR and address bar prediction in parallel. Their results can be combined to predict the texts in the address bar based on the coverage rate of text bounding boxes by address bar bounding box. 
For the above reasons, our final choice will be option O1.     

In either option, we need to correctly locate the address bar inside a photo, not only for accuracy reason, but more importantly for security reason. In particular, we deal with two types of domain name injection attacks. 
Figure~\ref{fig:ocr_example} shows the photograph of a simple webpage with three fields where a domain name may appear. Figure \ref{fig:ocr_example_result} is the corresponding OCR result. The first three lines include all the texts extracted from the photo and the fourth line represents the coordinates of the bounding box for all these texts. The following three lines show the texts and area locations of the title area, address bar area, and web page content area, respectively. Here OCR successfully extracts the texts and outputs their locations. However, it does not tell us which is the text from the address bar. As such, an adversary may want to get around PhotoAuth by injecting the valid domain name either in the title of the webpage or in its content.

To prevent the above injection attack, we need to make sure the correct domain is extracted from the OCR texts inside the predicted address bar area. Let $A_o$ and $A_b$ donate the area of a bounding box from OCR and that of a predicted addressbar, respectively.  Then we can define the metric \emph{cover rate} in the following formula, which basically indicates how much a text bounding box resulted from OCR overlaps with the predicted address bar. 
\begin{equation}
\begin{split}
Cover\ Rate (CR)=\frac{A_o\cap A_b}{A_o}
\end{split}
\nonumber
\end{equation}

\comment{
\vvv
\begin{equation}
\begin{split}
Cover\ Rate (CR)=\frac{OCR\ Result\ Area\cap Predicted\ Address\ Bar\ Area}{OCR\ Result\ Area}
\end{split}
\nonumber
\end{equation}
}

Only if CR is above a threshold value (to be determined in the evaluation section) and the extracted domain name matches with the server's, the user is accepted. In all the other cases, the user needs to retake a photo and try the second time. 

The other injection attack happens when an attacker embeds a fake browser address bar as an image inside the actual browser, the so-called ``picture-in-picture'' phishing attack~\cite{DBLP:conf/fc/JacksonSTB07} (or a similar case when the user has multiple browsers open).
As far as we are aware of, no real websites embed a second address bar in their login page, so it is certainly a very suspicious case. Therefore, in our system, whenever the server detects more than one address bar, it will reject the result and warn the user about phishing attack possibility. Meanwhile, it will request the user to retake a photo with only one address bar, which includes the website the user is actually visiting.    

\comment{
\begin{figure}[htbp!]
    \centering
      \begin{subfigure}[b]{0.225\textwidth}
        \includegraphics[width=\textwidth]{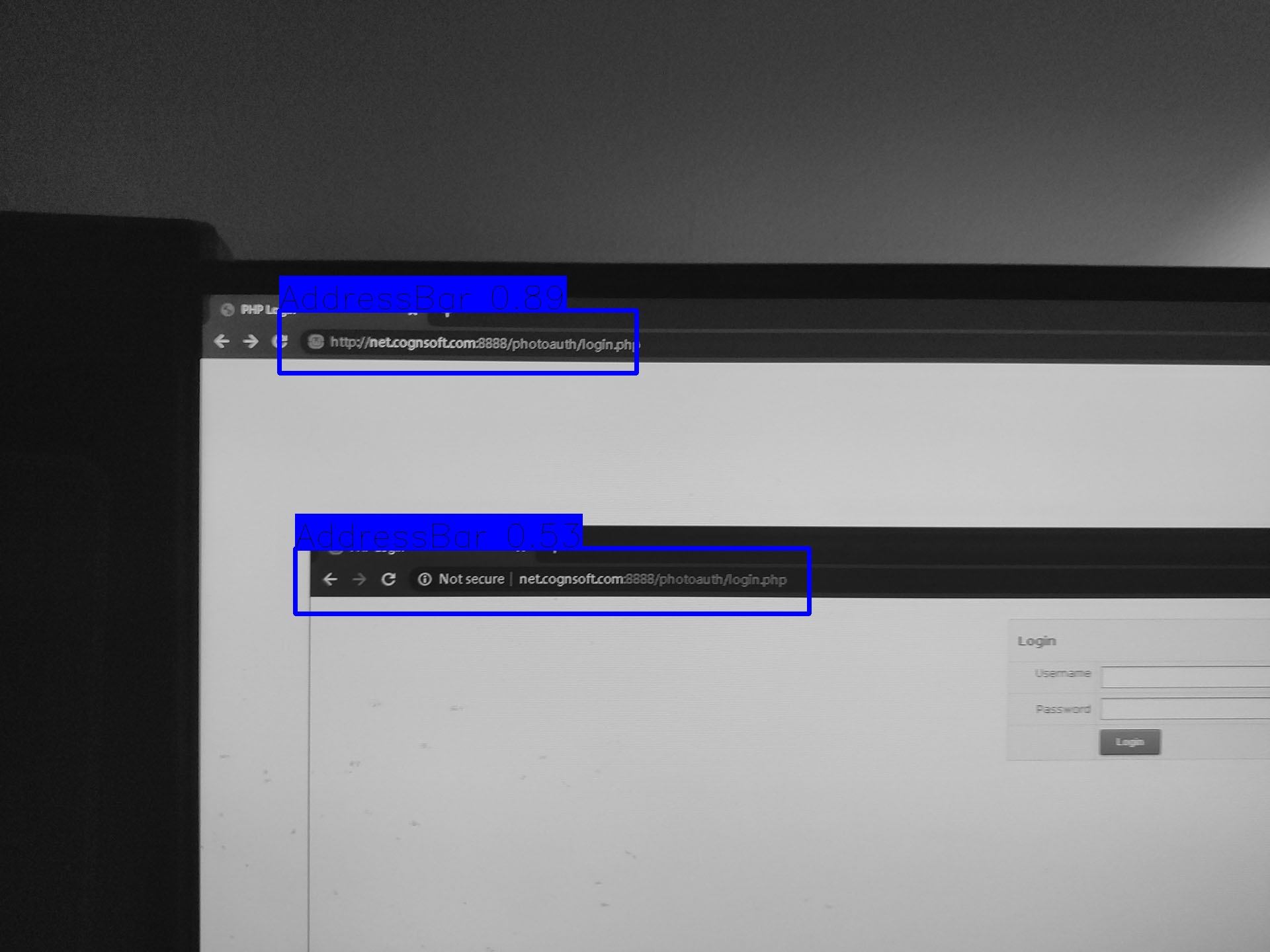}
          \caption{Picture in picture}
          \label{fig:picture_in_picture}
      \end{subfigure}
      \hfill
      \begin{subfigure}[b]{0.225\textwidth}
        \includegraphics[width=\textwidth]{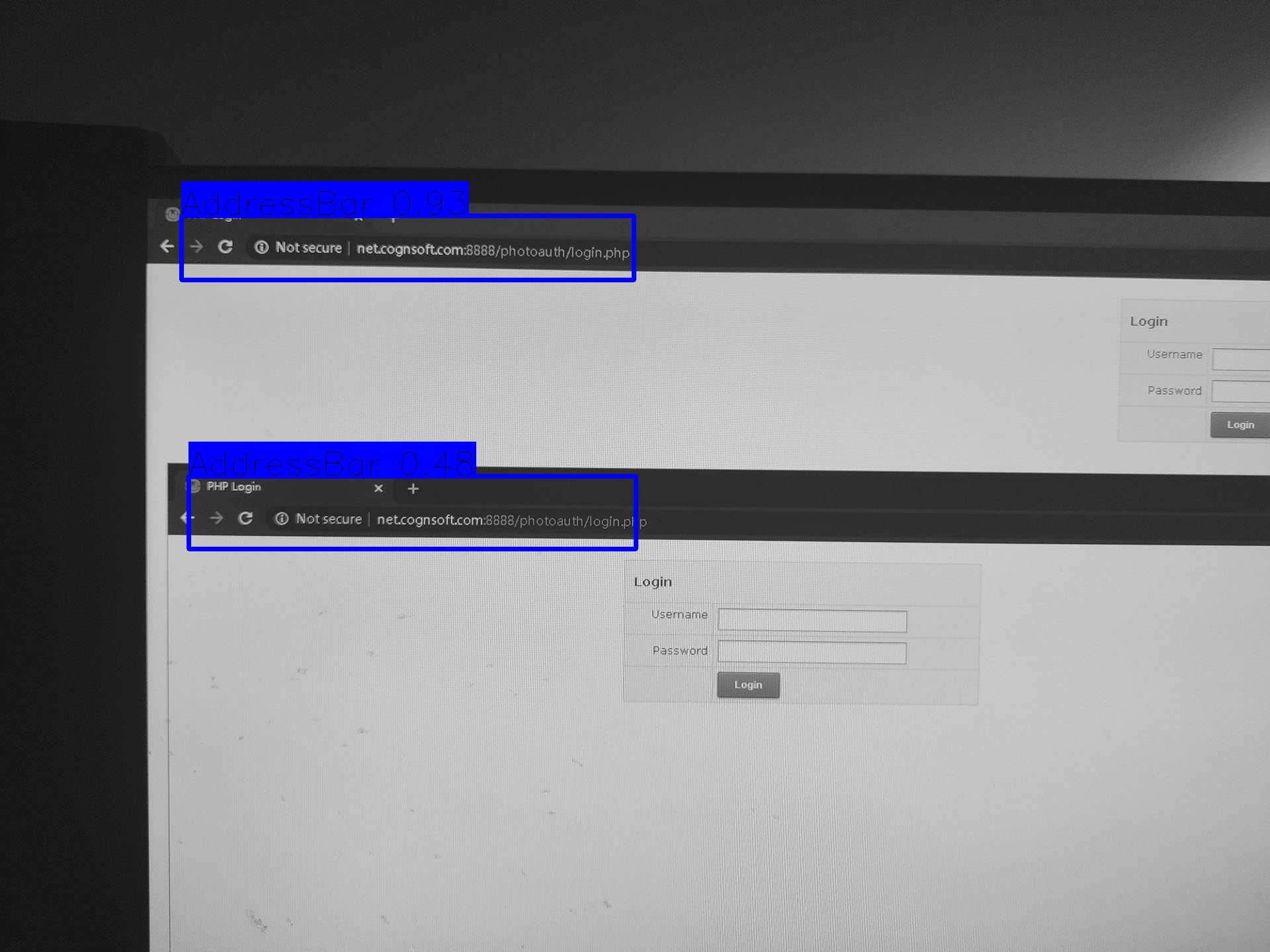}
          \caption{Two browsers}
          \label{fig:multiply_browser}
      \end{subfigure}

\caption{Multiple address bar examples}
\label{fig:multiple_address_bar}
\end{figure}
}

Currently, there is no tool for address bar identification, especially for address bars in photographed browser screens. 
Address bar identification is not a trivial task because users may take photos of \textit{different types} of browsers at \textit{different angles}, \textit{different distances}, and \textit{different illuminations}. Simply applying some heuristics based algorithms will not work well. Thus, we choose to train a deep learning based object detection model to predict the address bars .

\comment{

\subsection{Comparison with Other 2FA Methods}

\begin{table*}[]
\caption{PhotoAuth advantages over other 2FA methods}
\label{tab:advantage}
\begin{tabular}{l|l|l|l|l|l|l}
\hline
                                           & PhotoAuth & SMS OTP & Google Authenticator & Duo/Encap & RSA\_SecurID & U2F/Dongle \\ \hline
Based on unstable channel                  & No        & Yes     & No                         & No                      & No           & No         \\ \hline
Special device needed                      & No        & No      & No                         & No                      & Yes          & Yes        \\ \hline
Synchronize time                           & No        & No      & Yes                        & No                      & Yes          & Yes        \\ \hline
Waste computation when not use             & No        & No      & Yes                        & No                      & Yes          & No         \\ \hline
Physical read the code from distance       & No        & Yes     & Yes                        & No                      & Yes          & No         \\ \hline
Lost threat                                & No        & No      & No                         & No                      & Yes          & Yes        \\ \hline
Cannot erase 2FA  remotely                 & No        & Yes     & Yes                        & Yes                     & Yes          & Yes        \\ \hline
Compromised by interception                & No        & Yes     & Yes                        & No                      & Yes          & Yes        \\ \hline
Compromised by RTP                         & No        & Yes     & Yes                        & Yes                     & Yes          & No         \\ \hline
\end{tabular}
\end{table*}

Traditional 2FA has two types that are hardware token and software token. We evaluate different 2FA systems under the following features.
Based on an unstable channel: Does the 2FA based on an unstable communication channel that may have much delay?
Special device needed: Does the 2FA require a user to carry a special device every day just for authentication?
Synchronize time: Does system time matter? If time mess due to some application changing or system error, can the authentication process still work?
Waste computation when not use: When finish authentication, does the 2FA still need to compute extra things to wait for the next authentication.
Physical read the code from distance: If the victim closes the adversary, is there any way for the adversary the physical read the code \cite{raguram2011ispy,backes2008compromising,backes2009tempest} and input it before user input it?
Lost threat: If the 2FA device is lost, is there any protection mechanism for the 2FA device itself?
Cannot erase 2FA remotely: If the 2FA device is lost, is there any way to erase the 2FA system for fear the adversary use 2FA to authenticate?
Compromised by interception: Along with phishing, does 2FA also protect other attack methods like interception?
Compromised by RTP: Does 2FA compromised by real-time phishing?

\noindent \textbf{Software token}
SMS one-time password (OTP) is based on an SMS message. An SMS message is not very stable, especially for holidays like black Friday. It takes a while to receive an OTP SMS message.

Google Authenticator \cite{Google_2-Step_Verification,Google_Authenticator} generate OTP every 30 seconds and need to synchronize time with the server periodically. A user may only do authentication one time per day. All rest time still needs to generate OTP. This is a big waste. SMS OTP and Google Authenticator are both based on code, this is a potential threat, an adversary can read it in a distance\cite{raguram2011ispy,backes2008compromising,backes2009tempest}. 

Duo push \cite{Duo_Push} and Encap Security \cite{Encap_Security} are based on phone push message. A user needs to click the corresponding button to authorize or not. All of the software token solutions cannot solve the RTP problem.

\noindent \textbf{Hardware token}
RSA SecureID \cite{RSA_SecurID} is a special device generate OTP periodically. The only difference between RSA SecureID with google Authenticator is RSA SecureID runs on a special device. User needs to carry it every day just for authentication. RSA SecureID cannot solve the RTP problem.

Universal 2nd Factor (U2F) \cite{U2F} or Dongle \cite{Dongle} is a special anti-phishing device for authentication. U2F is the only current solution that can solve the RTP problem but with other problems.

Comparing with the current solution U2F, we have some advantages.

Our PhotoAuth is more convenient, everyone carries a phone today. Most people are reluctant to carry an additional U2F USB device every day just for authentication. Our PhotoAuth is not triggered for every authentication request. But with U2F, you have to push the OK button every time you authenticate in your account.

Our PhotoAuth is safer. U2F’s information still flow goes through the browser. Even data is encrypted, but this is still a potential threat especially when you use a public computer. Because a hacker may intercept your U2F flow information and give error information for the user to deceive the user that 2FA device is broken and the user should cancel 2FA this time to finish authentication. This is a potential threat. But in our PhotoAuth, all sensitive data go through the phone’s safe link. We exclude the interception threat. 

If an adversary first gets your password through phishing or check the leaked password database and then stolen your U2F. You are compromised. But with our solution, the adversary also needs to compromise the phone's fingerprint or facial biometric information. If your phone is lost, you can remotely erase your phone's 2FA. But U2F cannot do it. PhotoAuth gives one more step and raises the bar for the adversary.

As the table \ref{tab:advantage} shows, PhotoAuth can work under different channels. An unstable SMS channel delay is not a problem. PhotoAuth just needs a phone to install PhotoAuth application. No special device is needed. The authentication process of PhotoAuth does not need to synchronize time. PhotoAuth only costs resources on the authentication process, no need to periodically waste resources to waiting for the next authentication. PhotoAuth is not based on code, even an adversary read in distance will not harm. Even the phone is lost, the adversary has to first compromise phone password to gain access to the PhotoAuth. The user can remotely erase the 2FA to prevent adversary using. PhotoAuth's sensitive data all goes the phone's safe link, even the browser of the computer is intercepted, there is no way for an adversary to generate a fake message for the user to cancel 2FA for authentication. Last but not least, PhotoAuth will not be compromised by the RTP attack.
}

\section{Prototype Implementation}

\subsection{Environment}
On the server side, we choose Apache 2.4.39 as the web server, MySQL 5.7.26 as the database, and PHP 7.2.18 as the script language. The hardware configuration of the server is: CPU (I7-8700K Up to 4.7GHz), GPU (GTX 1080Ti 11GB GDDR5X), RAM (32GB 3200MHz) and it runs Windows 10. We use a Samsung S10 phone with 6GB RAM and Android 10.0.                                 
\comment{
The details of test device information can be found at Table \ref{tab:The_Server_and_User's_Phone_Devices_Detail}.

\begin{table}[h]
\caption{Specifications of Server and User's Phone Devices}
\begin{tabular}{l|l}
\hline
       & Server                                      \\ \hline
CPU    & I7-8700K Up to 4.7GHz                       \\ \hline
GPU    & GTX 1080Ti 11GB GDDR5X Boost Clock 1645 MHz \\ \hline
RAM    & 32GB 3200MHz                                \\ \hline
OS     & Windows 10                                  \\ \hline
       &                                             \\ \hline
       & User's Phone                                \\ \hline
Type   & Samsung S10                                 \\ \hline
RAM    & 6GB                                         \\ \hline
OS     & Android 10.0                                \\ \hline

\end{tabular}
\label{tab:The_Server_and_User's_Phone_Devices_Detail}
\end{table}
}
\subsection{Website Front-end Implementation}

\begin{figure}[htbp!]
    \centering
    \comment{
      \begin{subfigure}[b]{0.3\textwidth}
        \includegraphics[width=\textwidth]{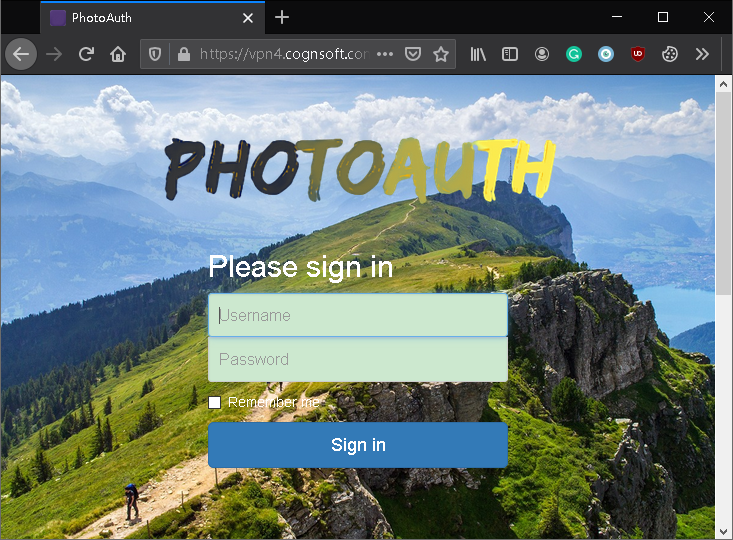}
          \caption{A Simple Website Interface}
          \label{fig:website_interface}
      \end{subfigure}
      }
      \begin{subfigure}[b]{0.2\textwidth}
        \includegraphics[width=\textwidth]{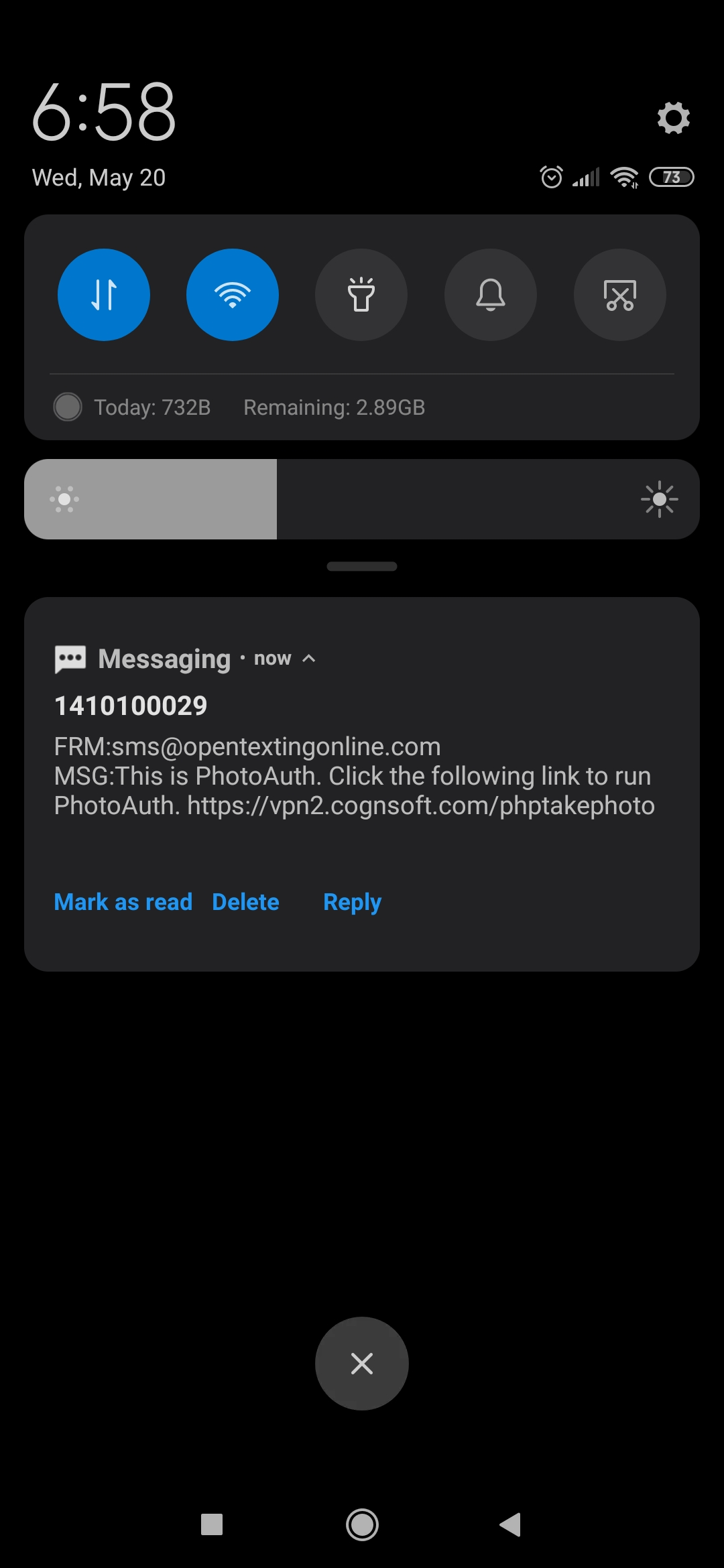}
          \caption{Phone receive PhotoAuth SMS message}
          \label{fig:application_interface_2}
      \end{subfigure}
      \hfill
      \begin{subfigure}[b]{0.2\textwidth}
        \includegraphics[width=\textwidth]{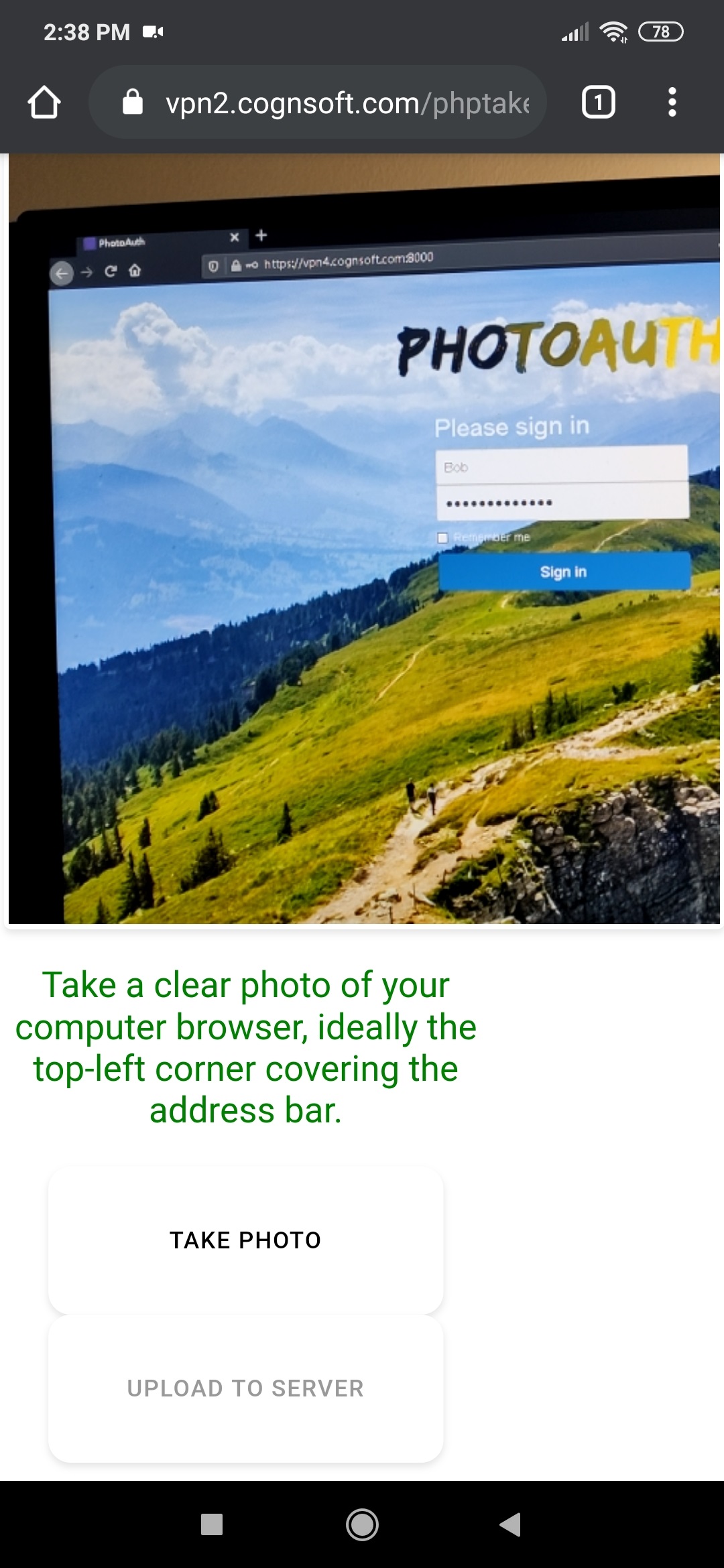}
          \caption{PhotoAuth web app GUI with a taken browser photo}
          \label{fig:app_interface}
      \end{subfigure}
\caption{Screenshots from our prototype implementation}
\label{fig:one_hash_represent}
\end{figure}

As a proof-of-concept prototype, we place two input boxes in the server login page to obtain the username and password. 
If both the username and the password are correct, 
the server passes the logic to the PhotoAuth module for 2FA.  
Then, the user's phone will receive a notification, which, in our prototype, is an SMS notification through a free online SMS provider~\cite{open_texting_online}, as shown in Figure \ref{fig:application_interface_2}. 
When the user taps the link inside of the SMS notification message, the default browser in the phone will load the PhotoAuth web app. 
We develop the PhotoAuth web app with HTML5+CSS+Javascript. The web app requires users to take a photo and then upload it. It will first request users to grant the browser camera-related permission for taking photos. The permission is requested \textit{only once} per website, and the browser makes sure the permission for the web server will not be misused by other websites. 
Given the permission, the web app will use ``navigator.mediaDevices'' to invoke the system camera app on the phone (Figure \ref{fig:app_interface}). Then the user takes a photo of the address bar in his PC browser by pressing the ``capture'' button in the camera app. 
The captured photo (in jpeg format) is displayed in the middle region of the GUI. If the user is happy with the photo, he can click the ``upload'' button to upload it to the server. 
We use the ``POST'' method of an asynchronous ``XMLHttpRequest'', a built-in browser object that allows to make HTTP requests in JavaScript, so that the image can be transferred to the server in the background. 
After the server receives the photo, it checks the domain name located in the address bar of the photo against its own one. If the domain name is correct, it will authorize the user to login; otherwise, it will deny the user.

\subsection{Back-end Implementation}
\label{sec:backend}

We use python to develop the server side back-end logic. Specifically, we use concurrent programming and multi-threading to achieve multi-tasking for better performance. 
For OCR, we directly use google vision API, mainly as a proof-of-concept. In practice, the server should deploy its local OCR software (e.g., the open-source PaddleOCR~\cite{paddleocr} for better privacy and efficiency). 
We use ``from google.cloud import vision" to upload a photo to the google vision server for OCR and get the result. 
For address bar detection, our main job here is to train a deep learning model. The question is: what are the important features of an address bar to differentiate it from any text-filled rectangle object? Our intuition is to leverage the commonly displayed icons, including the backward/forward arrows, reload and other icons on the left of an address bar.  Therefore, for training our model, we need to manually label the address bars with such surrounding areas as the ground truth. Manual labelling of address bars, however, is a time-consuming process. Therefore, for the proof-of-concept purpose,  
instead of labelling a huge dataset to train an entirely new deep learning model, we adopt transfer learning with a relatively small training set of 15,062 photos. 



Specifically, we choose Yolo v3 \cite{redmon2016you,redmon2018yolov3} as the object detection algorithm and adopt one of its pre-trained model~\cite{YoLo_v3_weights}. All its parameters and layers are kept as is. We use 50 epochs for transfer learning with learning rate 1e-3, batch size 32, and the Adam optimizer. In this stage, we only update the weights in the last three layers by freezing the weights in all the other layers. Then, in the next 50 epochs, we fine-tune the model with learning rate 1e-4, batch size 8 and Adam optimizer, while unfreezing all the layers to update all weights.  

As address bar detection is one type of object detection, we also use the Intersection over Union (IoU) metric to determine whether an address bar is detected correctly. IoU reflects how much the ground truth area and the predicted area overlap. During the training process (and also the testing process), IoU is an input parameter. If one sets the IoU score too big (for example above 0.8), only when the two areas fit very well with each other will it be considered a correct prediction; thus, the precision of the model will be very low. On the other hand, if the IoU score is too small (e.g. 0.2), then the predicted area might be too large (even cover some title areas). Therefore, in our model training, we use the default score 0.5, as used by other object detection models. 

\section{EVALUATION}

In PhotoAuth, authentication accuracy and scalability are very important performance metrics. 
Next, we evaluate bandwidth use, user non-action waiting time, PhotoAuth's OCR accuracy, address bar detection accuracy, and whole system accuracy.

\begin{figure*}[t!]
    \centering
        \includegraphics[width=0.8\textwidth]{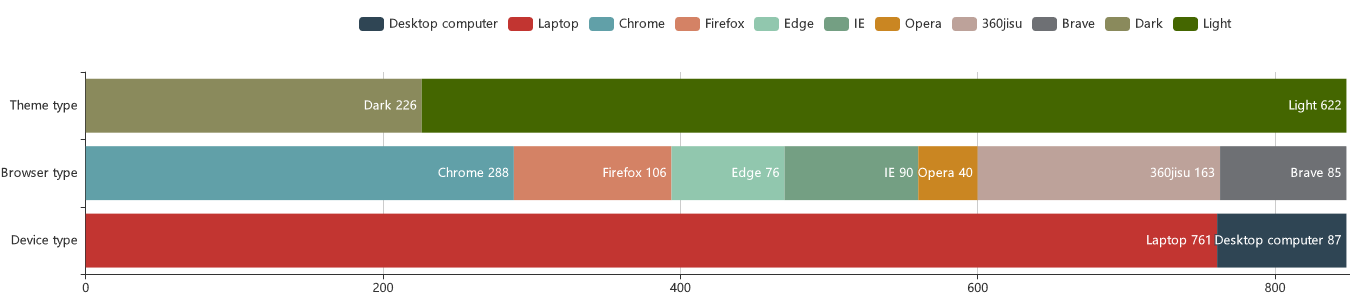}
          \caption{Test set composition}
          \label{fig:Composition}
\end{figure*}
      
\subsection{Dataset Composition}
\label{sec:dataset}

To train our transfer-learning based address bar detection algorithm, we took totally 16,454 photos.
Our data set is very diverse. 
First, there are two classes of devices initiating login requests, ``desktop", and ``laptop". ``Desktop" includes  monitors of different aspect ratio (e.g., 16:9, 21:9). 
``Laptop" includes different screen sizes (e.g., 14 inch, 15 inch).
Second, because different browsers have different fonts and different styles of address bars, our data set covers many famous browsers, including Chrome, Firefox, Edge, IE, Opera.
Third, there are numerous browser skins available (e.g., Chrome web store provides many themes), so it is too complicated to enumerate all browser skins. We chose two different windows color modes, ``light" and ``dark" themes. These themes change browser skins accordingly. Note that the PhotoAuth web app converts the colorful photos into gray-scale ones before uploading them to the server for OCR. In gray scale, skin personalization makes little difference from that of the light theme or dark theme. 

Fourth, the tilting angles and turning angles from the perfect shot angles are within the $[-15^{\circ},15^{\circ}]$ range and the shot distances are between 30-50 cm. Note that if we request users to take photos at a very close distance (e.g., 10 cm) and only cover the address bar region, the detection accuracy would be close to perfect. 
In our experiment, however, to make it more challenging, we took the photos from much farther away and the photos covered a very large portion of the PC screens, if not entirely. Furthermore, we also took photos in different environments and illumination settings.
Finally, all photos in our training and testing are resized into the resolutions of 1920x1080. 
1920x1080 resolution is supported by most cameras today. Therefore, if a user sets his camera at a higher resolution, the detection accuracy will be about the same because of the above resizing. 

Among 16,454 photos, we randomly picked 15,062 photos as the training set, 792 photos as the validation set, and 600 as one part of the test set. To test the transferability of our addressbar detection model, we then took 248 photos from 360jisu and brave browsers as the second part of the test set. Our trained model has not seen these two browsers before.
In the end, the test dataset contains 848 browser photos, which covers not only the the login page of our own test website but also 57 other popular websites (e.g., Chase, Bank of America). Figure~\ref{fig:Composition} shows the composition of our test dataset. 

\comment{
As mentioned in Section~\ref{sec:backend}, we set IoU threshold score as 0.5 in our training.  Figure~\ref{fig:addressbar_detection_example} and Figure~\ref{fig:URL_in_addressbar_CR_lower} show two examples. We then labelled the address bar areas in these photos. 

We constructed a dataset of PC browser photos for our evaluation, and 
Figure \ref{fig:Composition} shows its composition. 
To make the data set very diverse, we divide them into multiple classes. First, there are two classes of devices initiating login requests, ``desktop", and ``laptop". ``Desktop" includes  monitors of different aspect ratio (e.g., 16:9, 21:9). 
``Laptop" includes different screen sizes (e.g., 14 inch, 15 inch).
Second, because different browsers have different fonts and different styles of address bars, our data set covers many famous browsers, such as Chrome, Firefox, Edge, IE, Opera.
Indeed, as we will see the details in Section~\ref{sec:addressbar}, 
we trained our address bar detection model with the five types of browsers listed above. In addition, to test the transferability of the model, we will include in our test dataset 248 browser photos from two other types of browsers (360jisu and Brave).  

There are numerous browser skins available (e.g., Chrome web store provides many themes), so it is too complicated to enumerate all browser skins. In our test, we chose two different windows color modes, ``light" and ``dark" themes. These themes change browser skins accordingly. Note that the PhotoAuth web app converts the colorful photos into gray-scale ones before uploading them to the server for OCR. In gray scale, skin personalization makes little difference from that of the light theme or dark theme.


In the end, the test dataset contains 848 browser photos, which covers not only the the login page of our own test website but also 57 other famous websites (e.g., Chase, Bank of America).
All photos in our training and testing are resized into the resolutions of 1920x1080 irrespective of their original resolutions and the types of cameras used to take them. 1920x1080 resolution is supported by most cameras today. Therefore, if a user sets his camera at a higher resolution, the detection accuracy will be about the same because of the above resizing. 

}

\comment{
\begin{figure}[htbp!]
    \centering
      \begin{subfigure}[b]{0.45\textwidth}
        \includegraphics[width=\textwidth]{image/Composition.png}
          \caption{OCR test set composition}
          \label{fig:Composition}
      \end{subfigure}
      \hfill
      \begin{subfigure}[b]{0.4\textwidth}
        \includegraphics[width=\textwidth]{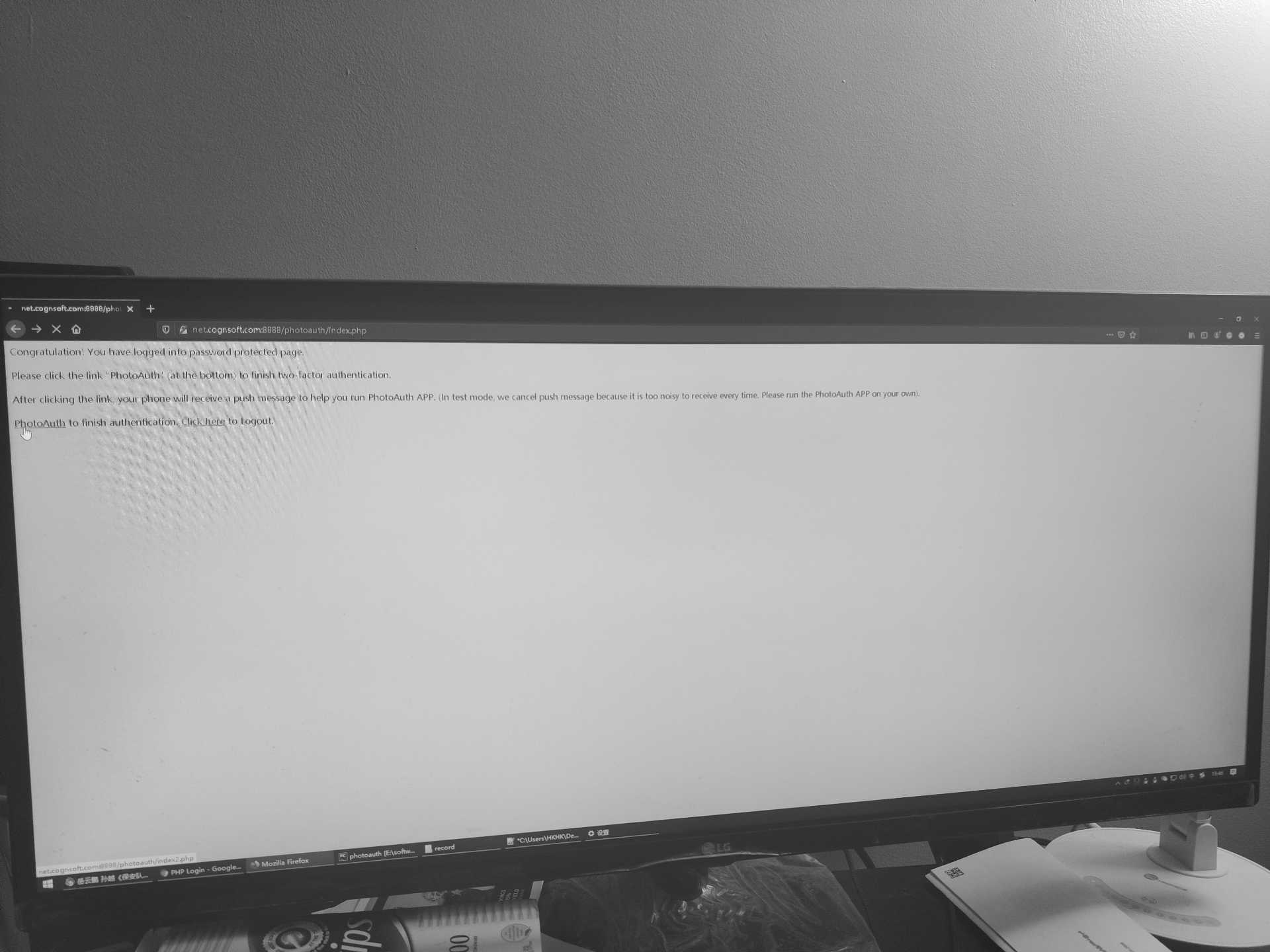}
          \caption{An example test case}
          \label{fig:test_set_example}
      \end{subfigure}
\caption{Test set}
\label{fig:test_set}
\end{figure}
}


\comment{
\subsection{Bandwidth Saving}
To save bandwidth for photo uploading, in our prototype we compressed the photos into small size, gray-scale ones. As OCR and objection detection algorithms do the same for their input, as long as the resolution of the compressed photos is good enough, it will not introduce errors into our system while saving bandwidth.   
The question is: what is the good photo resolution? To answer this question, we performed an empirical study by taking photos of monitor screens at different distances.
We found that at the capturing distance of about 1 meter, picture resolution starting from 1920x1440 (when camera aspect ratio is 4:3) or 1920x1080 (when camera aspect ratio is 16:9) up did not make difference in terms of OCR accuracy and address bar detection accuracy. In practice, people would most likely capture the photos of screens in a shorter distance, say within 50 cm, the typical distance between one's body and the PC. In this case, we found resolution of 800x600 is often sufficient. Anyway, to be more conservative, we recommend and adopt the 1920x1440 or 1920x1080 resolution in our study, depending on the camera aspect ratio.

\begin{figure}[h]
\centering
\includegraphics[scale=0.4]{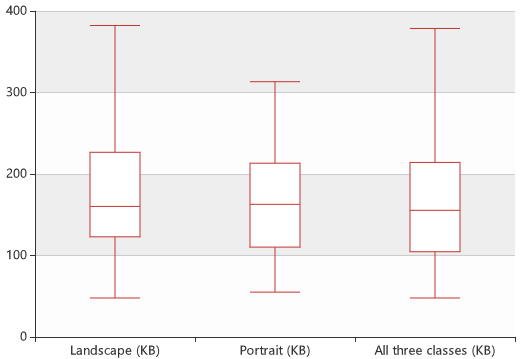}
\caption{Sizes of uploaded image files}
\label{fig:filesize}
\end{figure}

After taking a photo, the PhotoAuth web app resizes it while keeping its original aspect ratio. The final step is to convert the bitmap format into the jpeg format. We set the jpeg quality threshold as 50. A lower quality value means keeping less high-frequency information of the photo but saving more storage space. For example, we successfully compressed a raw bmp RGB photo (4000x3000) from 17.46 MB to a gray-scale jpeg photo (1920x1440) of size 160 KB. 

Figure \ref{fig:filesize} is the box plot of file sizes for 66 landscape photos and 70 portrait photos. 
The x-axis shows different types of image categories, which are landscape only, portrait only, and all mixed. Their median file sizes are 160.16 KB, 162.73 KB, and 155.35 KB, respectively. Landscape and portrait file sizes are almost the same.

\subsection{Bandwidth Use}
To save bandwidth for photo uploading, in our prototype the PhotoAuth web app compressed the photos into small size, gray-scale ones. 
Our empirical study showed that after compression, the output image resolution of 1920x1440 or 1920x1080 (depending on the camera aspect ratio) is good enough without effecting OCR detection accuracy or address bar prediction accuracy on the server side. The median image size generated in our experiment is around 160KB, which is small and hence fast to upload.  The details are presented in Appendix~\ref{sec:bandwidth}. 

}

\subsection{Bandwidth Use}
\label{sec:bandwidth}

To save bandwidth for photo uploading, in our prototype the PhotoAuth web app compresses the photos into small size, gray-scale ones. As OCR and objection detection algorithms do the same for their input, as long as the resolution of the compressed photos is good enough, it will not introduce errors into our system while saving bandwidth.   

The question is: what is the good photo resolution? To answer this question, we performed an empirical study by taking photos of monitor screens at different distances.
We found that at the capturing distance of about 1 meter, picture resolution starting from 1920x1440 (when camera aspect ratio is 4:3) or 1920x1080 (when camera aspect ratio is 16:9) up did not make difference in terms of OCR accuracy and address bar detection accuracy. In practice, people would most likely capture the photos of screens in a shorter distance, say within 50 cm, the typical distance between one's body and the PC. In this case, we found resolution of 800x600 is often sufficient. Anyway, to be more conservative, we recommend and adopt the 1920x1440 or 1920x1080 resolution in our study, depending on the camera aspect ratio. The size of a jpeg format file at this resolution is around 160KB in our study. 

\comment{
\begin{figure}[h]
\centering
\includegraphics[scale=0.5]{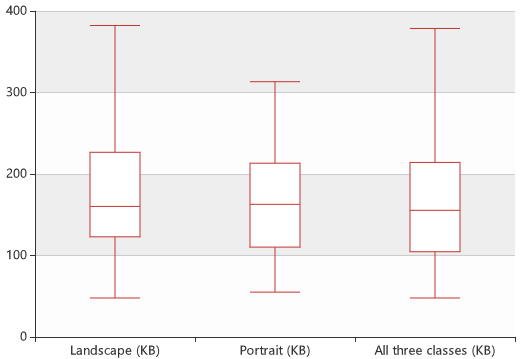}
\caption{Sizes of image files}
\label{fig:filesize}
\end{figure}

After taking a photo, the PhotoAuth web app resizes it while keeping its original aspect ratio. The final step is to convert the bitmap format into the jpeg format. We set the jpeg quality threshold as 50. A lower quality value means keeping less high-frequency information of the photo but saving more storage space. For example, we successfully compressed a raw bmp RGB photo (4000x3000) from 17.46 MB to a gray-scale jpeg photo (1920x1440) of size 160 KB.

Figure \ref{fig:filesize} is the box plot of file sizes for landscape photos and portrait photos. 
The x-axis shows different types of image categories, which are landscape only, portrait only, and all mixed. Their median file sizes are 160.16 KB, 162.73 KB, and 155.35 KB, respectively. Landscape and portrait file sizes are almost the same.
}

\subsection{User's Non-action Waiting Time}
To end users, the latency of our 2FA system is an important performance metric to care about, because it reflects the usability of our system. As users may spend different time to take photos, we will not consider the part of latency due to user's action. Instead, we test user's \textit{non-action waiting time}, which is mainly composed of photo uploading time and server side processing time. On the server side, OCR and address bar predication are carried out in parallel. Here we only count the OCR time, because our tests showed that it was always more than that for address bar prediction. We ignore the other types of waiting time in the system, because 
they are negligible compared to the two listed above. 

\comment{in Formula \ref{equ:non_user_action_waiting_time}. 

\begin{equation}
\begin{split}
User \ Non-action\ Waiting \ Time=\\
Photo \ Transmission \ Time+Photo \ Processing \ time\label{equ:non_user_action_waiting_time}
\end{split}
\end{equation}
}

Figure~\ref{fig:non_user_action_waiting_time} shows the uploading transmission time and overall non-action waiting time in both WiFi and LTE settings. The bandwidth of LTE is typically smaller than WiFi, so the transmission time of LTE (in our experiment the median is 1393 ms) is higher than that of WiFi (median 80 ms). Moreover, because in LTE many users share the same base station, so its transmission time fluctuates more than in WiFi case. The server side photo processing time does not depend on the uploading channel, either via LTE or via WiFi. In the WiFi setting, such processing time domains the overall waiting time, whereas in the LTE setting, the processing time accounts for roughly half of the waiting time.  

\begin{figure}[h]
    \centering
          \includegraphics[width=0.4\textwidth]{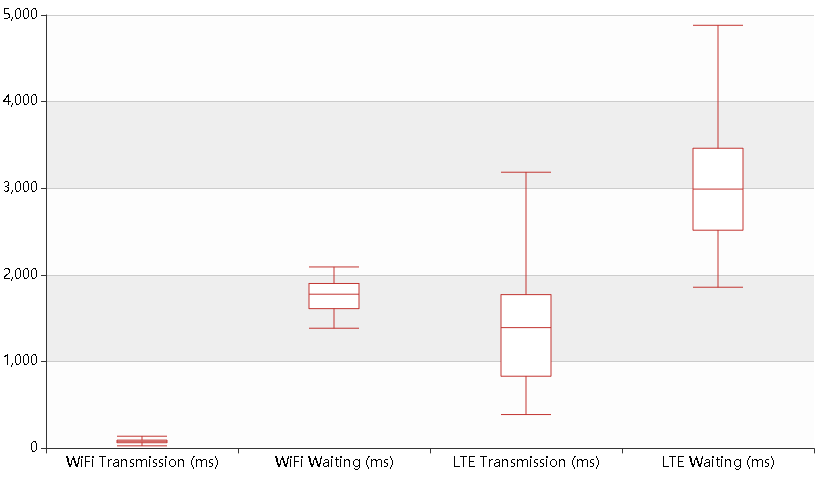}
          \caption{User's non-action waiting time}
           \label{fig:non_user_action_waiting_time}
\end{figure}
\comment{
\begin{figure}[h]
    \centering
      \begin{subfigure}[b]{0.45\textwidth}
        \includegraphics[width=\textwidth]{image/reaction_time.png}
          \caption{User's non-action waiting time}
           \label{fig:non_user_action_waiting_time}
      \end{subfigure}
      \hfill
      \begin{subfigure}[b]{0.45\textwidth}
        \includegraphics[width=\textwidth]{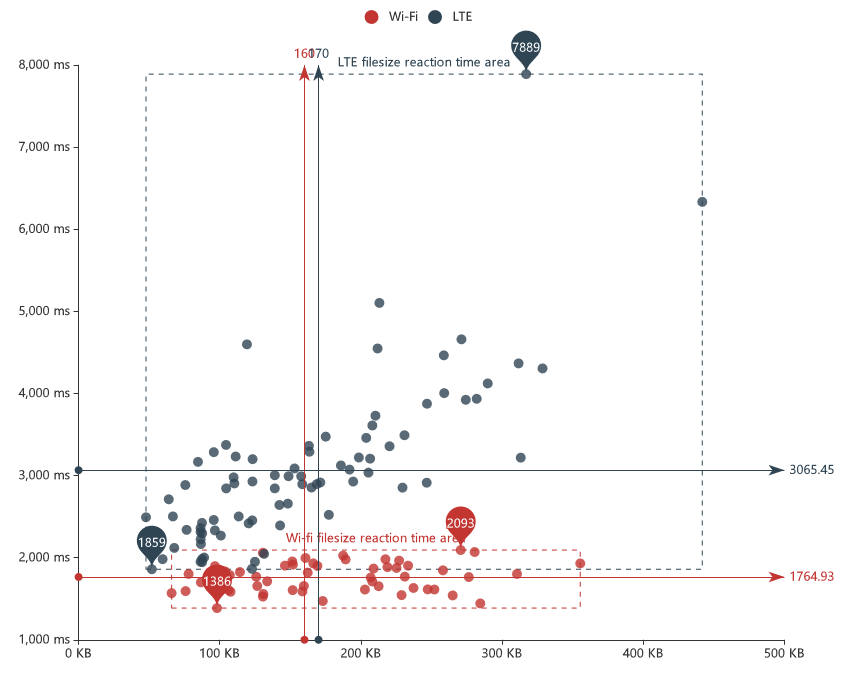}
          \caption{File size vs non-user action waiting time under WiFi and LTE}
           \label{fig:filesize_non_user_action_waiting_time}
      \end{subfigure}
\caption{User's non-action waiting time}
\label{fig:filesize_and_non_user_action_waiting_time}
\end{figure}
}

\subsection{Domain Name OCR Precision and Recall}
\label{sec:ocr}

PhotoAuth relies on accurate text output from OCR (specifically the texts from the address bar) to decide whether to authorize an authentication request. In this section, we test the precision and recall of OCR under various real-world settings to see how robust our system is to extract the domain names. 
Note that in the case when precision and recall are not perfect, it does not mean an adversary can bypass our system. We will elaborate on this point 
in Section~\ref{sec:whole_system_evaluation}.

In traditional object detection, an area is called a predicted area if the confidence score of detection is above a threshold. If the IOU (Intersection over Union) of the predicted area and the ground-truth area is above a specific threshold, it is a true positive; otherwise it is a false positive. If no area is predicted, it is a false negative. 

Domain name OCR is different from traditional object detection in that it has two stages: \textit{detection} and \textit{recognition}. A true positive occurs only when \textit{both} the right area (in our case, the address bar area) in an image is detected and the domain name (not the entire URL) inside the area is correctly recognized. Otherwise, we may have false negatives (when the area is not detected) or false positives (when the domain names are wrongly recognized).



\begin{figure}[htbp!]
    \centering
    
      \begin{subfigure}[b]{0.4\textwidth}
        \includegraphics[width=\textwidth]{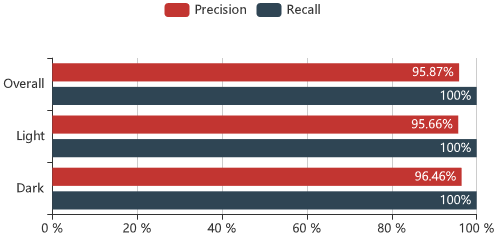}
          \caption{OCR precision and recall for both light and dark color modes and the overall}
          \label{fig:ocr_precision_and_recall}
      \end{subfigure}
      \hfill
      \begin{subfigure}[b]{0.4\textwidth}
        \includegraphics[width=\textwidth]{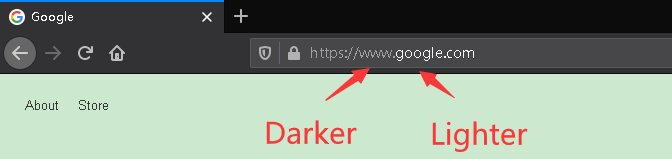}
          \caption{OCR false positive example}
          \label{fig:OCR_falsepositive_example}
      \end{subfigure}

\caption{OCR Experiment}
\label{fig:OCR_Experiment}
\end{figure}

Figure \ref{fig:ocr_precision_and_recall} shows the results with the 848 images (including 56 unique domains) introduced in our test set. 
The recall was 100\%. This means, as long as there was an address bar in the photo, OCR could detect the text area accurately. 
The OCR generated 35 false positives. There were two types of false positives. First,  the dot `.' in the domain names was too small to be recognized. Second, certain letters were mis-recognized, e.g., in one case `o' was recognized as `a' and the other case `o' as `e'.  The overall precision was 95.87\%.



Figure~\ref{fig:OCR_falsepositive_example} shows a failing example of OCR with the dark color mode of a browser. This is because the browser automatically made the ``www.'' part of a hostname (e.g., ``www.google.com'') darker, causing the OCR to miss `.'. Note that even in this challenging setting, this type of error only happened occasionally.   

Note that the precision of domain name OCR can be greatly improved with better quality photos. 
We believe that in practice, once users understand that the 2FA is based on the address bar content, they can naturally take photos at smaller distances while focusing on the address bars instead of the entire PC screen. 
We did an additional test with the domain names of Alexa top 50 websites~\cite{Alexarank}. We randomly changed 11 (out of 37) `o's into '0's, 5 (out of 17) `l's into `1's in the names. 
Differently, this time we took photos at the distance of about 20 cm and focused on the top-left corner. In the end, among the 527 characters in all these domains, there was only a single recognition error -- one `1' was recognized as `l'. The accuracy can improve further with a smaller distance.   





\subsection{Address Bar Detection Precision and Recall}
\label{sec:addressbar}

\begin{figure}[htbp!]
    \centering
         \begin{subfigure}[b]{0.45\textwidth}
        \includegraphics[width=\textwidth]{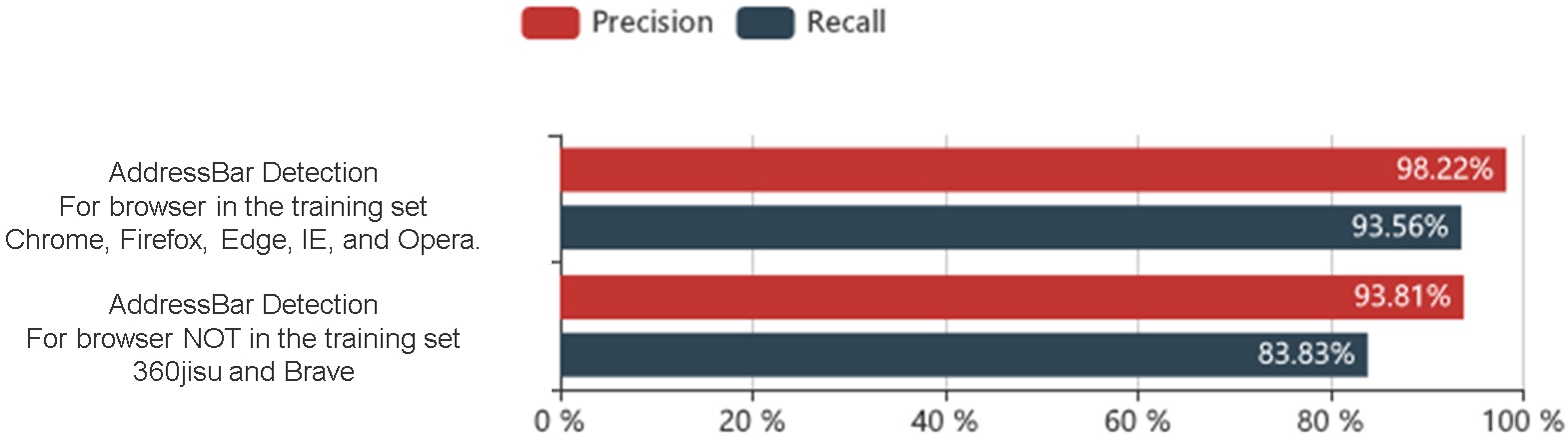}
          \caption{Address bar detection precision and recall.}
           \label{fig:addressbar_precision_recall}
      \end{subfigure}
            \hfill
      \begin{subfigure}[b]{0.35\textwidth}
        \includegraphics[width=\textwidth]{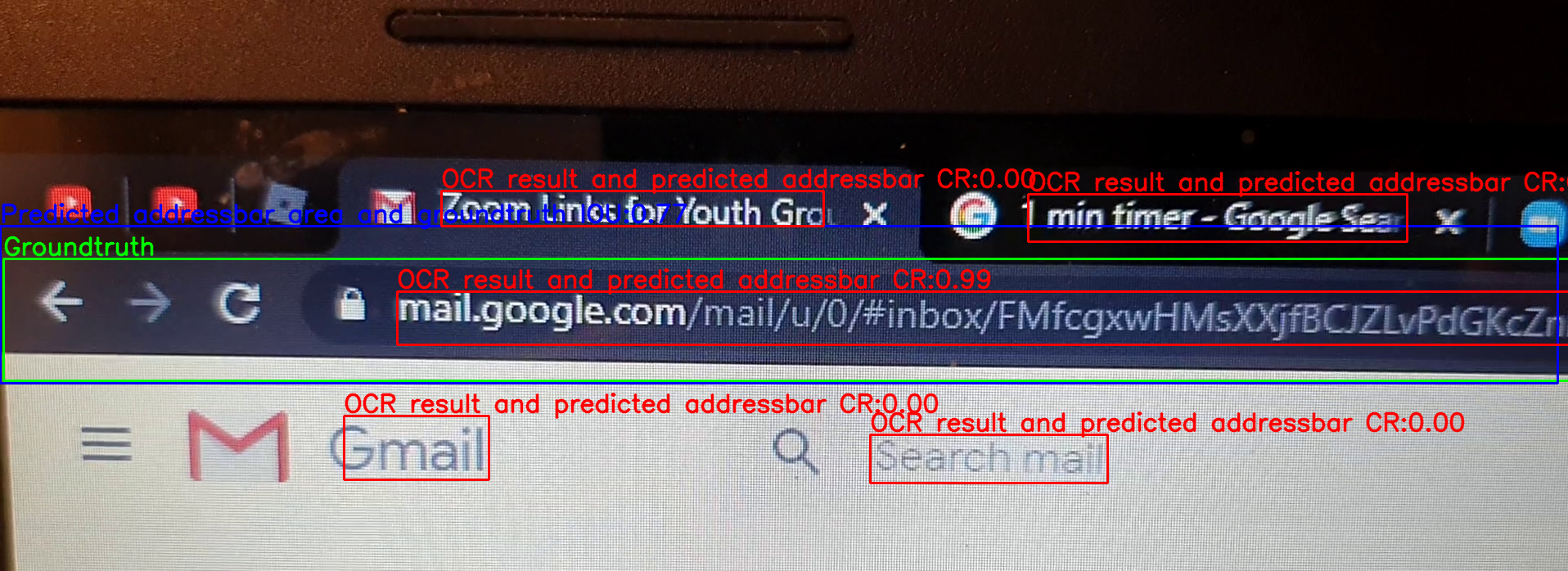}
          \caption{A true positive case}
          \label{fig:addressbar_detection_example}
      \end{subfigure}

\caption{Address bar detection}
\label{fig:address_bar_experiment_related}
\end{figure}

With the dataset introduced in Section~\ref{sec:dataset}, we train our addressbar detection model and measure its performance. 
Figure \ref{fig:addressbar_precision_recall} shows that the precision and recall of addressbar detection for known browsers (i.e., covered in the training set) are 98.22\% and 93.56\%, 
respectively. 
The precision and recall of addressbar detection for unknown browsers are 93.81\% and 83.83\%, respectively, which look reasonably good. This relatively lower accuracy is not surprising because the address bar features of Brave and 360jisu are different from that of the other five browsers. 
For wider deployment of our system, we believe a better approach is to train a model with additional types of browsers, for example, top 10 browsers. 

Figure~\ref{fig:addressbar_detection_example} shows an example output. The green bounding box is the labelled ground truth address bar area, and the blue one is the address bar area predicted by our deep learning model. One may notice that the ground truth area covers not only the address bar, but also commonly displayed icons on the left of the address bar as they provide the important features for address bar detection. 

\comment{
To train our transfer-learning based address bar detection algorithm, we took totally 16,454 photos of five major browsers (i.e., Chrome, Firefox, Edge, IE, Opera) under different environments, with variations on shooting angles, distances, illumination settings, PC types, browser types, as well as browser sizes (e.g., full-screen mode and windowed mode). As mentioned in Section~\ref{sec:backend}, we set IoU threshold score as 0.5 in our training.  Figure~\ref{fig:addressbar_detection_example} and Figure~\ref{fig:URL_in_addressbar_CR_lower} show two examples. We then labelled the address bar areas in these photos. Among 16,454 photos, we randomly picked 15,062 photos as the training set, 792 photos as the validation set, and 600 as one part of the test set (in Section~\ref{sec:dataset}). To test the transferability of our model, we then took 248 photos from 360jisu and brave browsers as the second part of the test set. }

As an object detection task, \textbf{not} an object classification task, our model either outputs a predicted address bar, or outputs nothing. It \textit{does not} know the ground truth area, although in our evaluation we manually labelled the ground truth areas to measure the detection accuracy. 
When no address bar is predicted in our testing, it is clearly a false negative. Now, when a address bar is predicted, it can be either a real one (IOU score above 0.5) or a false one (IOU score below 0.5).  In the example in Figure~\ref{fig:addressbar_detection_example}, the actual IOU is 0.77, which is above the threshold, so it is a true positive case.
If the actual IOU is under 0.5, which means the predicted area is much different from the ground truth area, it will be counted as a false positive case with respect to object detection. 

In practice, both false positive and false negative errors could cause the failure for the server to extract domain names correctly, and hence photo re-takes would be necessary. 
Also, in practice, our system may request users to take a photo that only focuses on the top-left corner of the browser at a closer distance. In this case, the address bar would be much easier to detect.  


\comment{
\begin{figure}[htbp!]
    \centering
      \begin{subfigure}[b]{0.225\textwidth}
        \includegraphics[width=\textwidth]{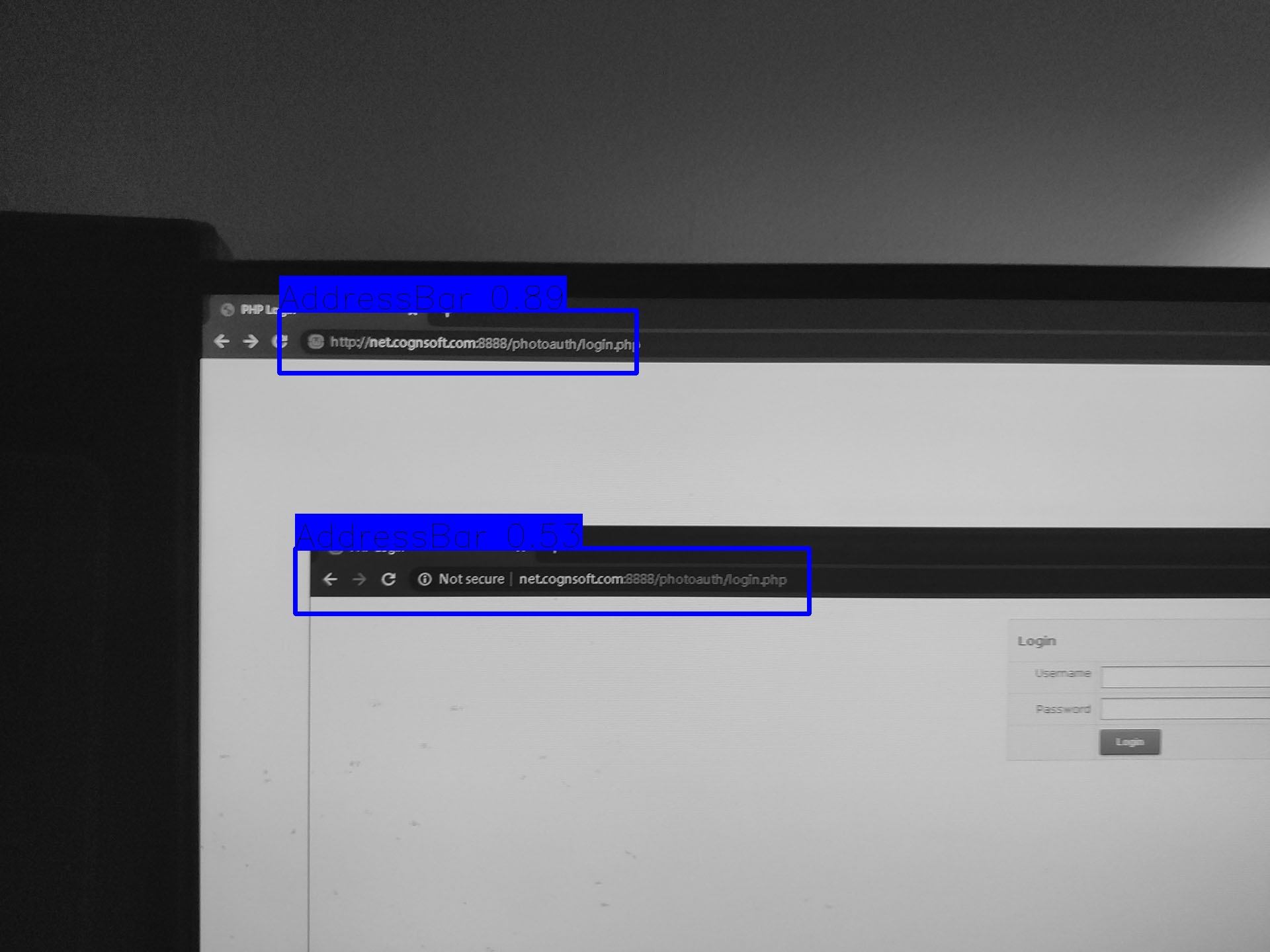}
          \caption{Picture in picture}
          \label{fig:picture_in_picture}
      \end{subfigure}
      \hfill
      \begin{subfigure}[b]{0.225\textwidth}
        \includegraphics[width=\textwidth]{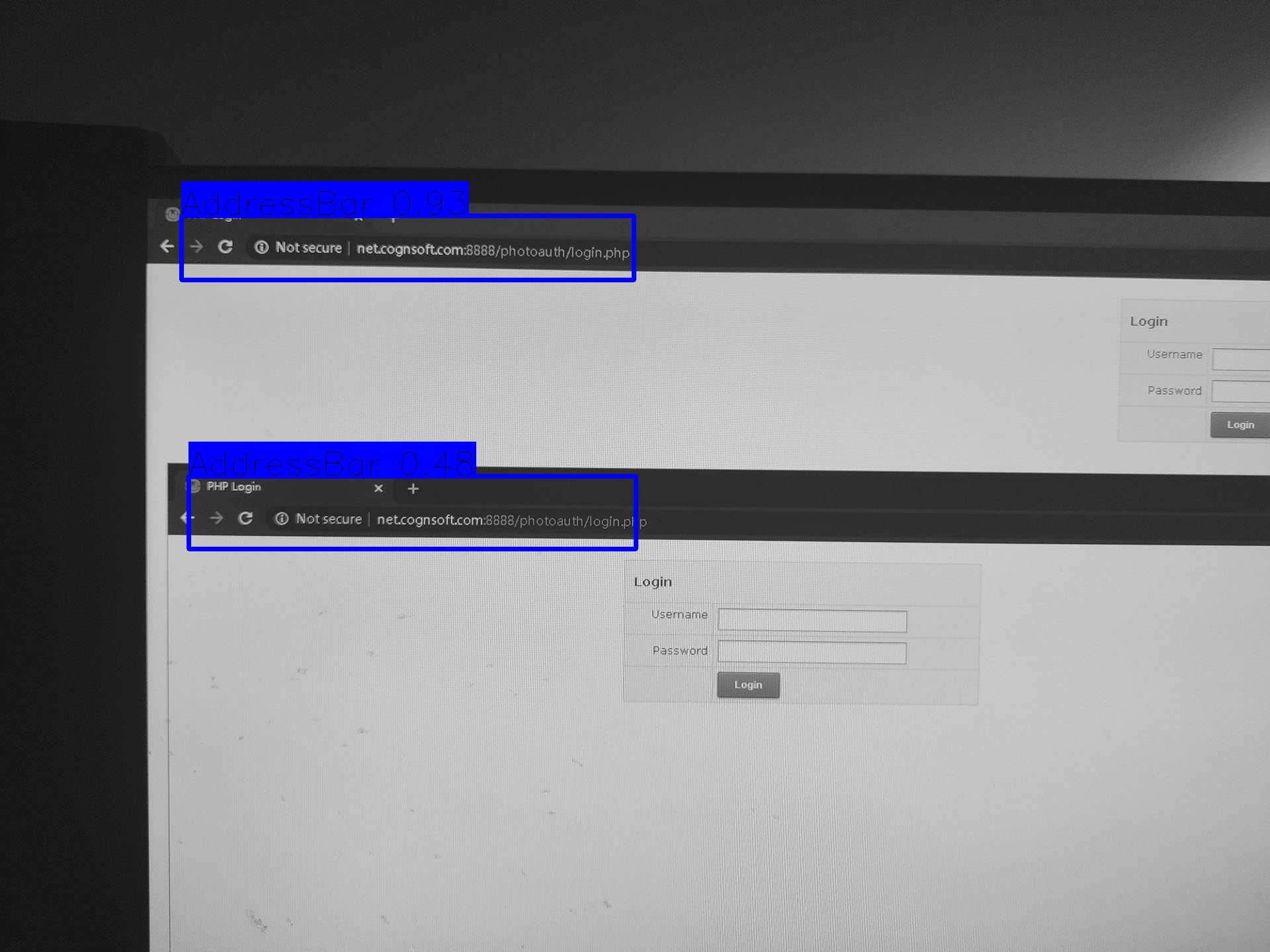}
          \caption{Multiply browser}
          \label{fig:multiply_browser}
      \end{subfigure}

\caption{Multiply address bar example}
\label{fig:multiply_address_bar}
\end{figure}
}

Based on the test dataset, we found that the median address bar detection time for one photo is 71 ms (the maximum 88 ms). It does not lag the whole system when parallelized with OCR because OCR takes at least 1 second to return the result. 




\subsection{Whole System Evaluation}
\label{sec:whole_system_evaluation}
\begin{figure}[h]
\centering
\includegraphics[width=0.35\textwidth]{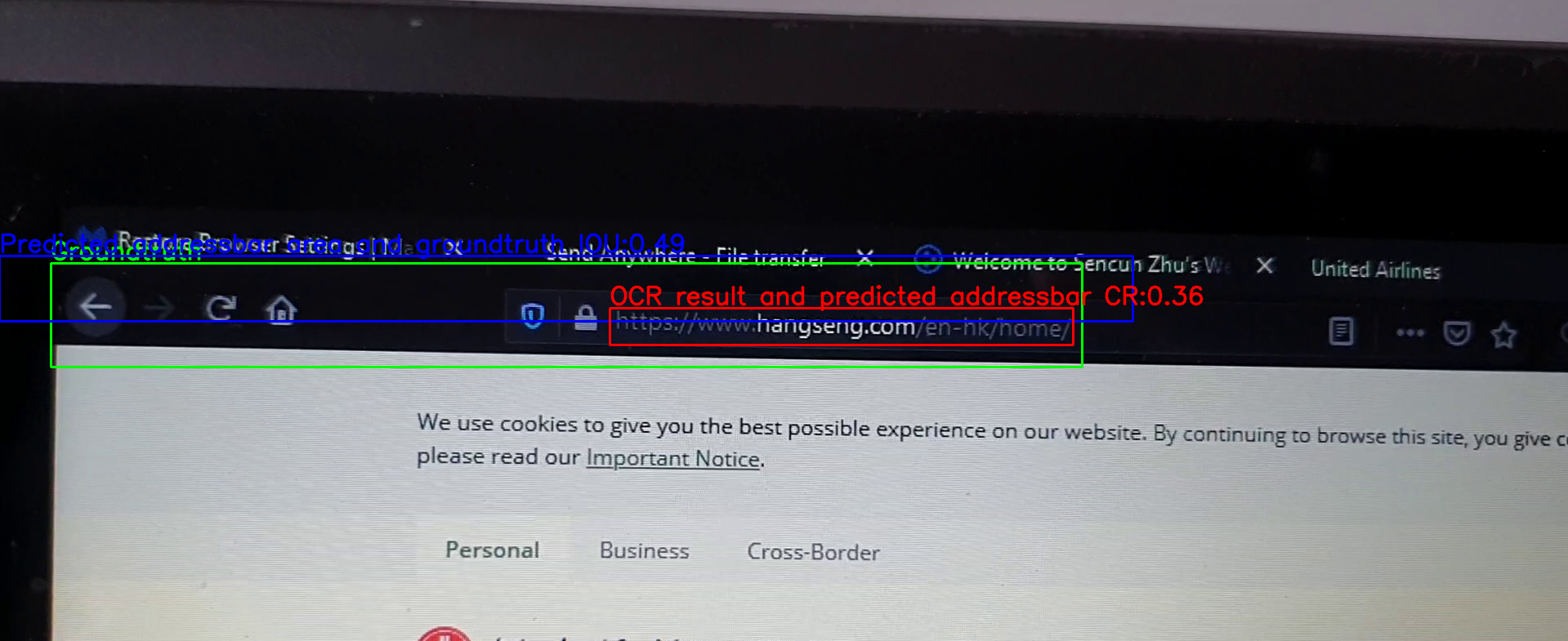}
          \caption{A case (CR=0.36) requiring a photo retake}
          \label{fig:URL_in_addressbar_CR_lower}
\end{figure}

\comment{
\begin{equation}
\begin{split}
Cover\ Rate (CR)=\frac{OCR\ Result\ Area\cap Predicted\ Address\ Bar\ Area}{OCR\ Result\ Area}
\end{split}
\nonumber
\end{equation}
}
In the whole system evaluation, we combine the detection results of OCR and address bar detection and report the final results.  
Figure~\ref{fig:addressbar_detection_example} shows an example with five bounding boxes for OCR texts, two for the texts in top titles, one for the URL in the address bar, and two others for texts in page content, they are all in red rectangles. Here we do not show the areas for texts extracted from the web page. The blue rectangle is the predicted address bar. For each red rectangle above, we calculated the cover rate (CR) (defined in Section~\ref{sec:server}), and got 0.32 (for the top-right title) and 0.99 (for the URL in the addressbar) and 0 for the rest. 
After analyzing all the data, we found that the CR threshold of 0.8 can distinguish texts in webpage content/title from texts in the address bar very well.  
Finally, we measure the error of our entire system with the metric named \textit{retake rate}. Despite the cause of errors, as long as the server failed to recognize the domain name correctly, in our measurement it was counted as a retake case, where the user is requested to take a photo again. 
We used the 600 photos in our test set to measure the retake rate while setting CR=0.8.
The retake rate was 6.83\%. 
It is relatively high mainly because the low quality of the photos (Figure~\ref{fig:URL_in_addressbar_CR_lower} shows an example), which introduced errors into OCR and address bar detection. In practice, a user can easily fix the problem by taking a photo at a closer distance, at the right angle and focusing on the top-left corner. 
In our testing, with the CR threshold of 0.8, there was not a single case where a title or any texts inside a webpage was mis-identified as a domain name. Only the address bar areas have been detected, which shows that the system worked as expected.  


\comment{
\begin{figure}
          \includegraphics[width=0.45\textwidth]{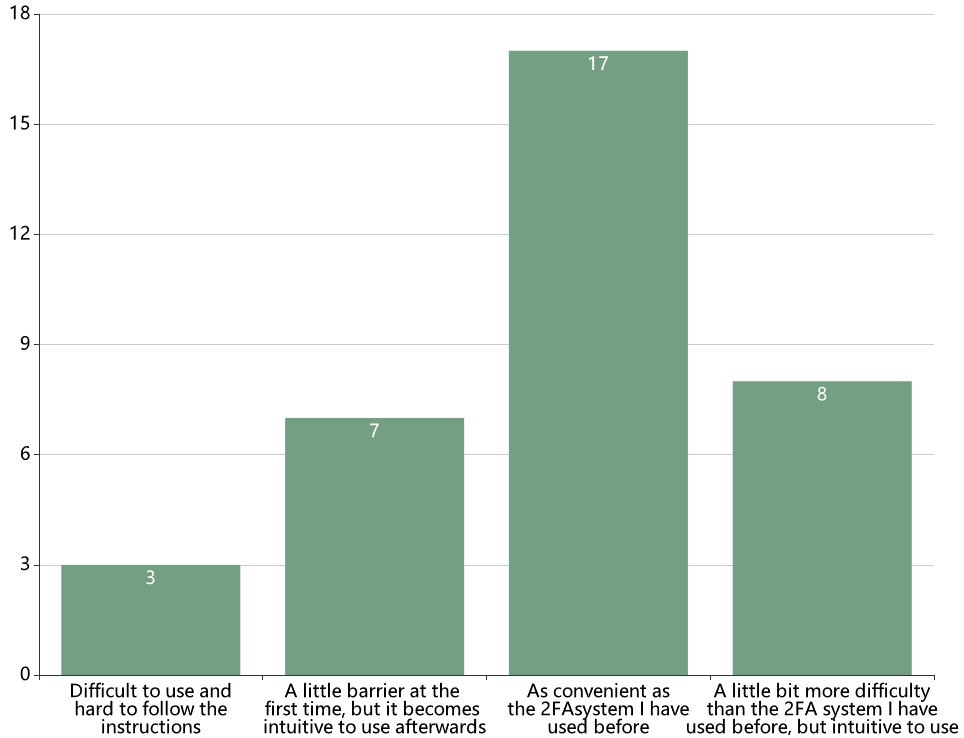}
          \caption{User's evaluation of PhotoAuth}
           \label{fig:rate_of_photoauth}
\end{figure}
}
Finally, we also conducted a preliminary user study over a demo version of PhotoAuth with 33 participants (IRB approved).  
The 33 participants showed a positive attitude on the usability f PhotoAuth (e.g., over 50\% considered it as convenient as the 2FA they have used before). Due to page limit, we present the details of our user study in the appendix for potential interest.

\section{DISCUSSIONS}
\label{sec:discuss}

\noindent \textbf{Evasion Attacks:}
An attacker may attempt to evade our system in different ways. First, 
as PhotoAuth relies on OCR to extract correct domain names, an attacker may register visually similar domain names (e.g., by registering g0og1e.com replacing  `o' with `0', `l' with `1', also called typosquatting domain names). This is one type of \textit{homograph attack}~\cite{homograph}. 
In Section~\ref{sec:ocr}, we have already shown that, with better-taken photos, the chance for this attack to succeed could be very low (1 error out of 527 characters in Alexa Top 50 domain names). As the OCR technique is advancing, such errors will be further reduced. Moreover, even if the attacker has successfully tricked a user into trusting his website (e.g., through phishing emails), 
he does not have the control over how the user takes photos; therefore, an OCR error may rarely result in a valid domain name, not to mention the case of perfectly turning into the target domain name. 
Moreover, the server may configure its web app to output a higher resolution photo instead of 1920x1080 
, as a tradeoff between communication time and OCR accuracy. 

  

\begin{figure}[!t]
\centering
\includegraphics[scale=0.45]{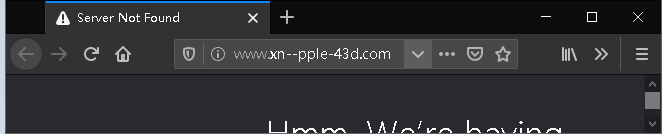}
\caption{The actual address bar content in Firefox when visiting ``аpple.com''.}
\label{fig:firefox_unicode}
\end{figure}

Another type of homograph attacks~\cite{DBLP:journals/cacm/GabrilovichG02,homograph} explore unicode for better success. For example, "apple.com" is different from "аpple.com". Even though they are the same looking, the letter "a" in the former one is ASCII (U+0041), whereas the letter "а" in the latter one is Unicode (U+0430). 
Not only a human user cannot recognize such Unicode letters easily, but the state-of-the-art OCR tools cannot recognize them correctly either. 
Fortunately, all major browsers 
only allow ASCII letters in the address bar, and 
they automatically convert the Unicode letters into the ASCII letters (Punycode). For example, Figure \ref{fig:firefox_unicode} shows that ``аpple.com” is  converted to ``http://xn--pple-43d.com” in the address bar. For this reason, such homograph attacks will not succeed. 

The second type of evasion attacks, which is specific to our system, is the 
domain name injection attack with multiple address bars. As we mentioned in Section~\ref{sec:server}, by design our system addresses this attack by requesting users to retake a photo containing only one address bar when the server detects more than one address bar in the photo.    
The third type of evasion attacks is a potential redirection attack against the workflow of our system. Specifically, after the user input his login credentials (Step 5 in Figure~\ref{fig:System_overview}), an attacker may try to redirect the user to microsoft.com, so that the user with PhotoAuth will take the picture on the genuine domain. However, this attack will not work. 
Recall that in Step 3, the real server (here ``microsoft.com'') sends a web cookie to the attacker. If the attacker replays (forwards) this cookie to user’s browser (Step 4), the user browser will store it for the fake webserver because it was received from the fake server instead of the real web server. Based on the same origin policy (SOP), the user’s browser will \textit{not} send this web cookie to the real web server if the user is instead redirected to the real web page that displays the same real login user interface. As a result, the real server will not receive a (valid) cookie from the user's browser, hence denying the login process. 



\comment{
\noindent \textbf{Increasing the Accuracy:}
We may leverage certain heuristics to reduce the OCR errors in our application context. For example, in the case of missing `.' after ``wwww'', we may infer the case based on the additional spacing between the last `w' and the character after it, which is introduced by the existence of `.'.  xxxxxx
}

\noindent \textbf{Handling Failures:}
When the web server determines that the detected domain name does not match with the real one, there is a small chance of detection errors. Since the error rate is very low in our system, we may set a threshold value (e.g., up to five) for the maximum number of retakes, meanwhile warning the user about potential phishing attacks and asking them to check the correctness of the domain name. Two factors may cause a legitimate user to fail in the 2FA process: poor photo quality (e.g., not focused, big angle, too dark) and poor network connection with the server. 
In either case, the user needs to redo the 2FA by taking a better photo or moving to a location with stronger wireless signal. The server can give the user some feedback on the cause, e.g., poor quality or timeout. For example, based on frequency information, with Fast Fourier Transform (FFT), it is easy to automatically detect whether an image is blurry~\cite{blurry}. 

Finally, just like many other 2FA systems, a user should register alternative authentication mechanisms (e.g, SMS one-time password), although they may provide weaker security guarantee. Certainly, caution has to be taken before allowing the system to fall back into a weaker mode. The user will need to be alerted about possible phishing attacks and check the running environment. 

\comment{
\noindent \textbf{Login on Phones:}
When a user (say Bob) logs into the web server (say microsoft.com) also with his mobile phone, we adopt a simpler strategy based on web cookies to defeat the RTP attack. Specifically, after he logs in with user name and password through his phone browser, the server sends back a web cookie to his browser. Meanwhile, 
the server sends a notification message to his phone (e.g., through PUSH or SMS). Similar to the process in Step 8 of our system (Section~\ref{systemoverview}), the message contains a short link like ``microsoft.com/c/689527'', where the 6-digit number is randomly generated as Bob's session id. This session id and the web cookie is linked for this specific login. 
Now, when Bob clicks the short link on his phone, his phone browser will load the web page ``microsoft.com/c/689527'',  which hosts a simple web app with JavaScript code to retrieve the previously deposited web cookie (permitted based on the same-origin policy). If the cookie is correct, the authentication will pass; otherwise, it will fail. Under a RTP attack, the attacker will not know the short link for submitting the cookie, so the web server (here microsoft.com) will not be able to (quickly) receive the correct cookie from the phone, so it can easily defeat the attack. Moreover, we can increase the length of the random number to defeat brute force attacks.    

Here an alternative idea could be applying PhotoAuth in a similar way by requesting a user to take a screenshot of his mobile browser (instead of taking a photograph of his PC browser). Since screenshots can always guarantee the picture quality without worrying about various factors (e.g., lighting, angle), it seems to be a good idea. However, there are two drawbacks. First, mobile browsers do not have APIs to support screenshotting. Therefore, the web app will not be able to take a screenshot with a button in the web page. While users can manually take a screenshot of his phone at any time by pressing certain hardware buttons in combination, there is no convenient way to pass the screenshot to the web app without additional steps. Second, due to the small screen size, the mobile browser may not be able to display the full domain name, making the system vulnerable to URL-truncation based phishing attacks~\cite{DBLP:conf/nsdi/NiuHC08,DBLP:conf/ccs/LuoSHN17}  that use a very long sub-domain name to match the full domain name of a victim site.  
Due to the above reasons, we recommend our web-cookie based design for mobile logins, which is simpler but more secure.  

\noindent \textbf{Alternative Implementations:}
In our system, we assume the phone link between user's smartphone and the website is secure from interception (e.g., man-in-the-middle attack) or interruption attacks by the adversary, no matter it is through SMS, email link or push message.  
In our implementation, we used SMS as the channel to pass the notification messages from the server to the client phone. One advantage is that it can be directly upgraded from the existing common SMS-based 2FA methods. Another advantage is that it is platform independent. That is, no matter which OS the client phone is using, either Android or iOS, our web-based PhotoAuth runs the same. The implementation would be almost the same if we use emails to pass the link.  

An alternative is to use push messages. In this case, the server provides its own mobile app (TLS/SSL protected) for users to install and each user registers an account and logs into his account before receiving push messages. While this method may provide a better protection of the phone link than SMS and email do, it is platform dependent. That is, the server needs to provide different mobile apps for different phone operating systems. Moreover, as one of our design goals is to offer high usability which does not require users download and install any additional software on there phones, we did not choose this alternative in our prototype. However, for applications where such a usability issue is not a concern, our system can be easily adapted to use push messages. The main difference is that the mobile app will receive notification messages instead.      
}

\noindent \textbf{Possible Limitations:} To users who have never used any 2FA system before or even do not know what browser address bar is, they will need to first spend a few minutes to understand the workflow of PhotoAuth (e.g., watching a short tutorial video, as provided in our user study) before logging into a PhotoAuth-enabled server. Otherwise, they may fall into various phishing attacks made possible through social engineering or other types of human errors. Clearly, it requires a joint effort from multiple parties (e.g., web server, web browsers, ISPs) to protect all users, both tech-savvy and non-tech-savvy, from such online phishing.

\comment{
\noindent \textbf{Difference with Password Managers} 
Both password managers (PM) (e.g, LastPass, 1password) and PhotoAuth are based on domain names. 
PM decides which pair of username and password to fill in edit boxes of a login page based on its domain name. 

Because the password manager needs to check the domain. So the password manager will not fill in username and password to a phishing website. When a user visits a phishing website, a user may treat the fake site as a real one. But at the same time, the password manager will not fill with credentials because of the fake domain. The password manager is based on analyzing domain and HTML structure like figure \ref{fig:password_manager} shows, it based on properties of ``input id", ``class", ``type", and ``name" to decide which edit box to fill. 

If the password manager does not fill in credentials, is that means the website is phishing site? The answer is no. If a property of the edit box is changed, it fails in filling username and password (e.g, website changes its webpage layout). For example, if the website change the ``input id" from ``password" to ``pwd", this causes a password manager to fail in filling credentials. In this case, a user may treat a real site as a phishing site. If such a webpage layout changing frequently. Even a user accessing a truly phishing site and the password manager is not filling any credential, the user may think the website just changing its layout again, the user thinks I just need to re-input my credentials to this site again. In such a case, the user is compromised. So password manager filling or not can not be a criterion for a phishing site.

\begin{figure}[!t]
\centering
\includegraphics[scale=0.4]{image/password_manager.png}
\caption{An example of how the password manager fills the form.}
\label{fig:password_manager}
\end{figure}

One more thing, a password manager is an option for a user. Not all users would install a password manager in their browsers. After installing a password manager, you have to trust this third party extension to maintain all your credentials. It is inconvenient for a user to use a password manager in a public computer. For example, you have to install a third-party password manager APP on a public computer every time. Sometimes, a public computer is not allowed to install any extension. 

PhotoAuth is a compulsory authentication mechanism rather than an option for a user. As long as the server detects abnormal signals of a user’s network environment, the server will challenge the user with the PhotoAuth. The whole process does not need any extension support from the browser (No third party extension or a plugin for the browser is needed.). PhotoAuth can compatible with almost all legacy devices.
}

\comment{
\begin{figure}[!t]
\centering
\includegraphics[scale=0.1]{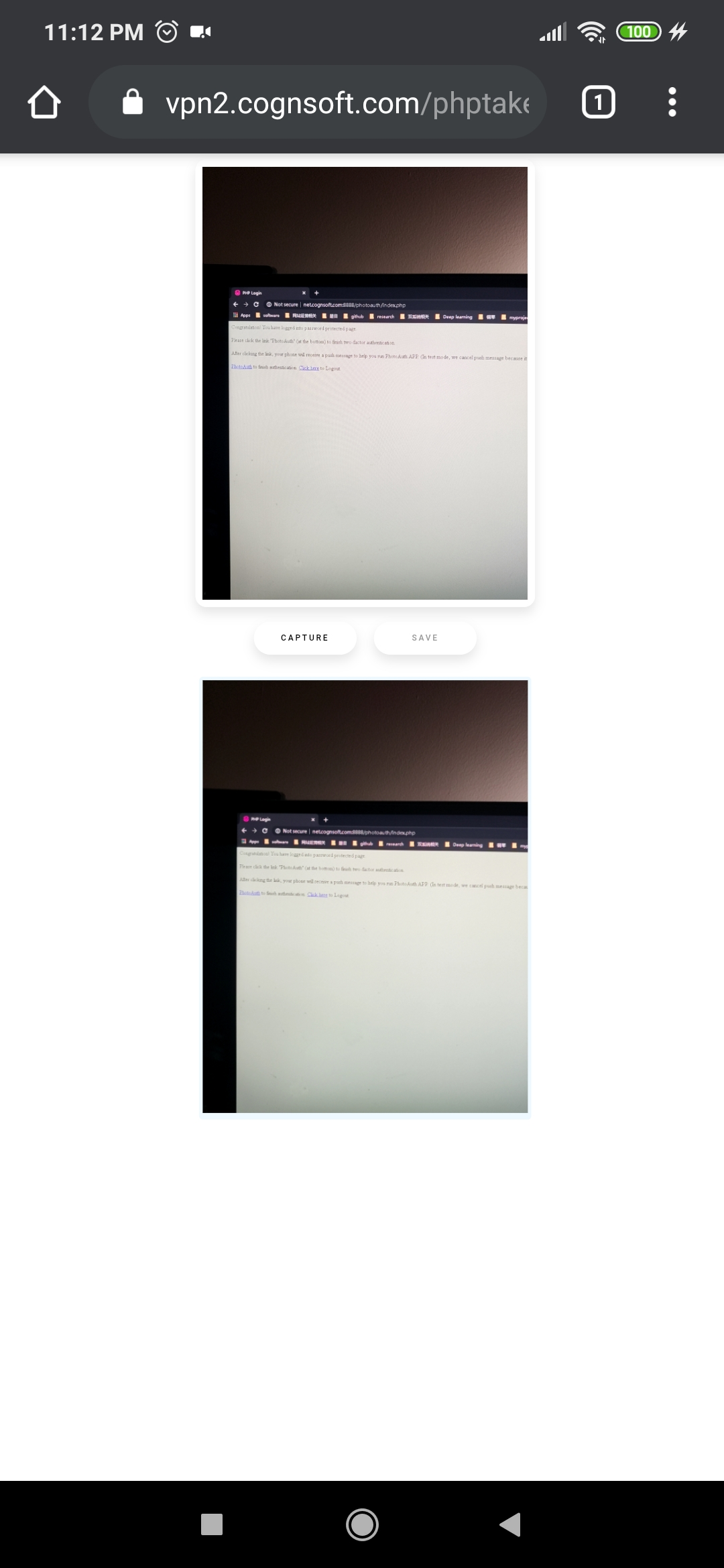}
\caption{Web based version of PhotoAuth}
\label{fig:phptakephoto}
\end{figure}
}

\section{RELATED WORK}

\subsection{Industrial Solutions} \label{sec:whyfail}
Google 2-Step Verification \cite{Google_2-Step_Verification} is a phone application to generate a time-based one-time code for the user in every 30 seconds. No network connection is required between the app and the server. The mechanism requires the user to manually enter the one-time password into the browser. 
Duo Push \cite{Duo_Push} is also a phone application that receives push information when the user sends a request in the browser login page and the user taps a button to respond. 
Unfortunately, they 
are vulnerable to the RTP attacks. 
In the 2nd factor authentication phase, the adversary can deceive the user to pass his one-time password or press the ``Approve'' button in the above two applications. The user may think the 2FA is for himself to authenticate to the website, but the truth is that the 2FA is for the adversary to get authentication. 

Recently Google released a new software-based 2FA tool leveraging phone's built-in security key \cite{Google_phone's_built-in_security_key}. It requires pre-installed phone app to generate the key, special built-in browser support, and Bluetooth (or NFC) to establish a secure channel between the computer and the phone,
such requirements could restrict its usability. 
In 2017 the FIDO Alliance proposed a Universal 2nd Factor (U2F) protocol~\cite{U2F}, where end users carry a single U2F device which works with any relying party supporting the protocol. Later, the FIDO Alliance proposed FIDO 2~\cite{Fido2}, by integrating its  Client-to-Authenticator Protocol (CTAP) with W3C's Web Authentication (WebAuthn). 
Users may log into internet accounts using their preferred devices. Web services and apps can turn on this functionality via biometrics, mobile devices and/or FIDO security keys. U2F/FIDO2, based on public-key cryptography, can counter RTP attacks very well. 

While the industrial solutions look promising to solve the RTP attacks, it may take a long time to be widely deployed because of 
several possible factors such as 
cost and usability issues~\cite{slow2adopt}. 
For instance,  
U2F devices are not free, commonly ranging from 20 to 60 dollars, hence a non-trivial cost overhead for either end users or companies. 
Other use options may require pairing between phone and PC, or BlueTooth or NFC. The concepts and procedure for deploying U2F/FIDO2 could still look complicated to some non-tech-savvy users because of the needed registration, installation or configuration. 
To this end, alternative secure and user-friendly solutions are still very needed. 
  


\comment{
\noindent \textbf{1FA for Traditional Phishing Detection}
There are several One Factor Authentication methods for detecting the traditional phishing by comparing the web page element similarity (e.g, image size, position) or site layout~\cite{rosiello2007layout,medvet2008visual}.
Unfortunately, such methods would not work in real-time phishing, because the fake websites keep updating with the victim websites.
Aggarwal et al., ~\cite{aggarwal2012phishari} proposed a real-time phishing site detection system based on domain name checking.
This method also fails when an adversary changes its domain frequently like every one hour. Other  research \cite{teraguchi2004client,herzberg2004trustbar} proposed adding a new security toolbar to detect phishing. A later study~\cite{wu2006security} shows that additional security toolbar cannot perfectly resist phishing. 
}

\subsection{Academic 2FA Solutions} 
Dhamija et al., ~\cite{dhamija2005battle,dhamija2005phish} proposed Dynamic Security Skins (DSS) to allow the server to prove its identity based on visual hash generated by the browser and the server. 
It has two major weaknesses: relying on users to determine genuineness, and incapable of preventing the RTP attacks. 
Shirvanian et al., \cite{shirvanian2014two} proposed a 2FA system based on mix-bandwidth devices. 
Even though the system can improve the usability of 2FA, it cannot be widely implemented because the requirements are not met by most devices. 
Czeskis et al., ~\cite{czeskis2012strengthening} proposed a 2FA system named PhoneAuth. Its overall protocol shares the same same spirit with U2F protocol except it involves a smartphone instead of a USB dongle. 
Parno et al., ~\cite{parno2006phoolproof} proposed a system to establish a secure bookmark on the phone side to control the authentication. 
Azimpourkivi et al.~\cite{azimpourkivi2017camera} introduced a camera-based 2FA system called Pixie, which 
establishes trust between a user and his web server based on both the knowledge and possession of an an arbitrary physical object.
However, the lack of a binding between the trinket and the website the user is visiting leaves the system vulnerable to RTP attacks.


Karapanos et al., ~\cite{karapanos2015sound} proposed a 2FA system based on the ambient sound. Basically, both user's browser and user's phone record the ambient sound at the same time. If the sound signals are much different, it is likely an attack case. The system can handles RTP attacks with the support of Bluetooth and microphone recording.
Ulqinaku et al.~\cite{DBLP:conf/wisec/UlqinakuLC19}  proposed 2FA-PP, which leverages a Web Bluetooth API, 
to create a secure Bluetooth connection between a website and the user’s smartphone that runs a special mobile app. 
To defeat phishing attacks, it leverages network latency measurements to tell if the user is connected to the legitimate server or to the attacker’s site. The system has high accuracy when the attackers are not located within the same region as the victim.







\section{CONCLUSION}
In this paper, we proposed PhotoAuth, a 2FA system to defend against real-time phishing (RTP) attacks. In PhotoAuth, a user takes a photo of the PC browser with the address bar area, and uploads the photo to the server. The server automatically extracts the domain name information from the address bar and detects fake domain names. 
PhotoAuth is easy to use and also compatible with the traditional 2FA system to support most legacy devices. 
It does not require special hardware (except user's phone),
We prototyped the system and tested it in various environment settings and with multiple types of browsers. The results showed that PhotoAuth is able to effectively prevent and detect attacks. 

\bibliographystyle{IEEEtranS}
\bibliography{references}
\section{APPENDIX}

\subsection{A Preliminary User Study}\label{userstudy}

\begin{figure}[htbp!]
    \centering
    
    
      \begin{subfigure}[b]{0.3\textwidth}
        \includegraphics[width=\textwidth]{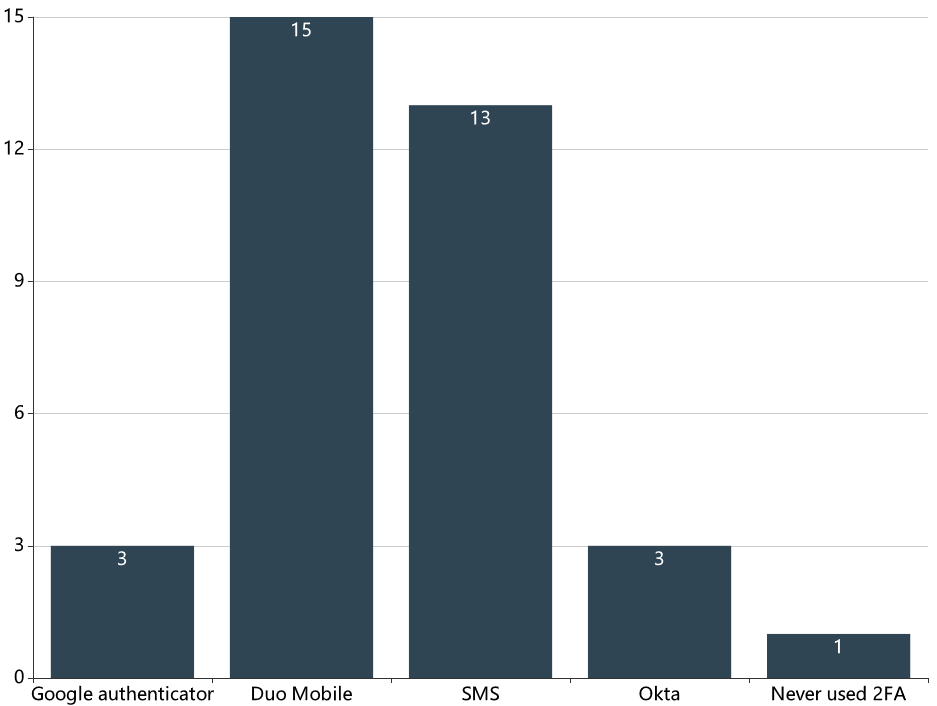}
          \caption{User's prior experience with 2FA}
          \label{fig:background_of_2fa}
      \end{subfigure}
      \hfill

      \begin{subfigure}[b]{0.3\textwidth}
        \includegraphics[width=\textwidth]{image/rate_of_photoauth.png}
          \caption{User's evaluation of PhotoAuth}
           \label{fig:rate_of_photoauth}
      \end{subfigure}
      
\caption{A Preliminary User study}
\label{fig:user_study}
\end{figure}

A comprehensive user study is intractable at this stage of our research for various reasons. Ideally, we may set up a working system for each participant to 
have the first-hand experience. However, in this case a user's opinion about our system may be greatly influenced by the details of our implementation, such as the performance of the system at the specific use time, the design of the web GUI, etc. Indeed, in our prototype we use a free online service for delivering SMS messages to mobile phones. Such a service is not stable (in one case it took five minutes to deliver) and occasionally SMS messages may even be blocked by cellular service providers. Too many differences between our prototype system and a real-world commercial system that would deploy PhotoAuth will prevent us from obtaining meaningful user feedback.    

As such, in this stage, we only conduct a preliminary user study regarding the overall workflow of the system. 
We made a short live tutorial video (around 1 minute 30 seconds) to demonstrate the 2FA process step-by-step. It starts with the username/password login page, 
followed by several UIs in both PC and phone (including those UIs in Figure~\ref{fig:one_hash_represent}). Each UI in the video has one or two sentences subtitle (instead of voice) to explain the high-level idea or procedure. As a result, users do not know exactly how our technique works through this video tutorial. We choose to do so because our study cares more about the overall usability of the system for ordinary users. After that, a user is asked to answer a questionnaire with three simple questions.
We obtained the IRB approval from our university for this anonymous user study, which does not ask and collect any personal information from participants.

We tried to recruit participants from our personal social networks with various  technical background. As the survey was anonymous and the participation was totally voluntary, we do not know who submitted the survey and do not know the demographic information of the participants. 
In the end, we collected 33 survey feedback.   


Figure~\ref{fig:background_of_2fa} shows that most participators have used a 2FA system before, with SMS and Duo Mobile being the most popular, so they have the basic understanding of 2FA. Figure~\ref{fig:rate_of_photoauth} shows that slightly over 50\% of users considered PhotoAuth ``as convenient as the 2FA system I have used before", 24.2\% users rated it ``a little bit more difficult than the 2FA system I have used before but intuitive to use'', 21.2\% of users rated ``a little barrier at the first time, but it becomes intuitive to use afterward''. Most users held a positive attitude regarding the usability of PhotoAuth. 

Clearly, this is only a preliminary user study. In our future work, we will consider designing and conducting a more comprehensive user study.  



\subsection{Alternative Implementations} \label{altimp}

\noindent \textbf{Handling the Case of Phone Login}
When a user (say Bob) logs into the web server (say microsoft.com) also with his mobile phone, we may simplify PhotoAuth and adopt a simpler strategy based on web cookies to defeat the RTP attack. Specifically, after he logs in with user name and password through his phone browser, the server sends back a web cookie to his browser. Meanwhile, 
the server sends a notification message to his phone (e.g., through PUSH or SMS). Similar to the process in Step 8 of our system (Section~\ref{systemoverview}), the message contains a short link like ``microsoft.com/c/6895272013'', where the 10-digit number is randomly generated as Bob's session id. This session id and the web cookie is linked for this specific login. 
Now, when Bob clicks the short link on his phone, his phone browser will load the web page ``microsoft.com/c/6895272013'',  which hosts a simple web app with JavaScript code to retrieve the previously deposited web cookie (permitted based on the same-origin policy). If the cookie is correct, the authentication will pass; otherwise, it will fail. Under a RTP attack, the attacker will not know the short link for submitting the cookie, so the web server (here microsoft.com) will not be able to (quickly) receive the correct cookie from the phone, so it can easily defeat the attack. Moreover, we can increase the length of the random number to defeat brute force attacks.    

Here an alternative idea could be applying PhotoAuth in a similar way by requesting a user to take a screenshot of his mobile browser (instead of taking a photograph of his PC browser). Since screenshots can always guarantee the picture quality without worrying about various factors (e.g., lighting, angle), it seems to be a good idea. However, there are two drawbacks. First, mobile browsers do not have APIs to support screenshotting. Therefore, the web app will not be able to take a screenshot with a button in the web page. While users can manually take a screenshot of his phone at any time by pressing certain hardware buttons in combination, there is no convenient way to pass the screenshot to the web app without additional steps. Second, due to the small screen size, the mobile browser may not be able to display the full domain name, making the system vulnerable to URL-truncation based phishing attacks~\cite{DBLP:conf/nsdi/NiuHC08,DBLP:conf/ccs/LuoSHN17}  that use a very long sub-domain name to match the full domain name of a victim site.  
Due to the above reasons, we recommend our web-cookie based design for mobile logins, which is simpler but more secure.  

\noindent \textbf{More Secure Implementation Choices:}
In our system, we assume the phone link between user's smartphone and the website is secure from interception (e.g., man-in-the-middle attack) or interruption attacks by the adversary, no matter it is through SMS, email link or push message.  
In our implementation, we used SMS as the channel to pass the notification messages from the server to the client phone. One advantage is that it can be directly upgraded from the existing common SMS-based 2FA methods. Another advantage is that it is platform independent. That is, no matter which OS the client phone is using, either Android or iOS, our web-based PhotoAuth runs the same. The implementation would be almost the same if we use emails to pass the link.  

An alternative is to use push messages. In this case, the server provides its own mobile app (TLS/SSL protected) for users to install and each user registers an account and logs into his account before receiving push messages. While this method may provide a better protection of the phone link than SMS and email do, it is platform dependent. That is, the server needs to provide different mobile apps for different phone operating systems. Moreover, as one of our design goals is to offer high usability which does not require users download and install any additional software on there phones, we did not choose this alternative in our prototype. However, for applications where such a usability issue is not a concern, our system can be easily adapted to use push messages. The main difference is that the mobile app will receive notification messages instead.      



%



\end{document}